\PassOptionsToPackage{noend}{algorithmic}
\documentclass[lettersize,journal]{IEEEtran}

\usepackage{glossaries}

\usepackage{amsmath,amsfonts}
\usepackage{orcidlink}
\usepackage[caption=false]{subfig}
\usepackage{stfloats}
\usepackage{url}
\usepackage{verbatim}
\def\BibTeX{{\rm B\kern-.05em{\sc i\kern-.025em b}\kern-.08em
				T\kern-.1667em\lower.7ex\hbox{E}\kern-.125emX}}
\usepackage{balance}

\IEEEoverridecommandlockouts
\usepackage{amssymb}
\usepackage{graphicx}
\usepackage{textcomp}
\usepackage{xcolor}
\usepackage{pgfplots}
\usepackage{bm}
\usepackage{array}
\usepackage{tikz}
\usepackage{mathtools} %
\usepackage{comment}    %
\usetikzlibrary{fit}
\usetikzlibrary{calc, shapes.gates.logic.US, shapes.gates.logic.IEC}
\usetikzlibrary{shapes.geometric, arrows.meta, calc, positioning, fit}
\usepgfplotslibrary{groupplots,dateplot}
\usetikzlibrary{positioning}
\usetikzlibrary{spy,backgrounds}
\usetikzlibrary{shapes.multipart}
\usetikzlibrary{positioning,calc}
\usepgfplotslibrary{fillbetween}
\usetikzlibrary{patterns}
\colorlet{mydarkblue}{blue!30!black}
\usepackage{multirow}  
\usepackage{placeins} %
\usepackage{soul}   %
\usepackage{algorithm}
\usepackage{algpseudocode}

\algrenewcommand\algorithmicindent{0.7em}%
\algnewcommand{\IfThenElse}[3]{%
	\State \algorithmicif\ #1\ \algorithmicthen\ #2\ \algorithmicelse\ #3}

\pgfplotsset{compat=1.16}
\definecolor{darkblue}{rgb}{0,0.2706,0.541}
\definecolor{darkgreen}{rgb}{0.098,0.4784,0.5176}
\definecolor{lightgreen}{rgb}{0.3412,0.7411,0.7647}
\definecolor{orange}{rgb}{0.9255,0.4313,0}
\definecolor{darkred}{rgb}{0.7529,0,0}
\definecolor{black}{rgb}{0,0,0}
\def\BibTeX{{\rm B\kern-.05em{\sc i\kern-.025em b}\kern-.08em
    T\kern-.1667em\lower.7ex\hbox{E}\kern-.125emX}}

\usepackage{standalone}
\usepackage{nicefrac}

\definecolor{darkgray176}{RGB}{176,176,176}
\definecolor{goldenrod1911910}{RGB}{191,191,0}
\definecolor{lightgray204}{RGB}{204,204,204}
\definecolor{darkgray176}{RGB}{176,176,176}
\definecolor{darkviolet1910191}{RGB}{191,0,191}
\definecolor{green01270}{RGB}{0,127,0}
\definecolor{lightgray204}{RGB}{204,204,204}

\definecolor{myorange}{RGB}{236,110,0}

\definecolor{coloribx}{RGB}{0,69,138}
\definecolor{coloriby}{RGB}{192,0,0}
\definecolor{coloribt}{RGB}{25,122,132}

\newcommand\myrv[1]{\mathsf{\MakeUppercase{#1}}}
\newcommand\mylabel[1]{\mathsf{#1}}
\newcommand\myopexpecation{E}%

\newcommand{\mathbbR}{\mathbb{R}} %
\newcommand\myvec{\boldsymbol}%
\newcommand\mytuple{\bm}%
\newcommand\myset{\mathcal}

\newcommand\mytranspose{\top} %
\newcommand{\breal}{b}
\newcommand{\hatb}{\hat{b}}
\newcommand{\hatl}{\hat{\ell}}
\newcommand{\labelcn}{\mathsf{c}}
\newcommand{\labelvn}{\mathsf{v}}
\newcommand{\labelch}{\mathsf{ch}}
\newcommand{\myK}{K}
\newcommand{\kb}{k_\mathsf{b}}

\newcommand{\Nt}{N_\mathsf{t}}
\newcommand{\Z}{Z}
\newcommand{\z}{z}
\newcommand{\setZ}{\mathcal{Z}}

\newcommand{\iby}{y}
\newcommand{\ibY}{\myrv{Y}}
\newcommand{\ibyalp}{\mathcal{Y}}
\newcommand{\ibxreal}{x}
\newcommand{\ibx}{x}
\newcommand{\ibX}{\myrv{X}}
\newcommand{\ibxalp}{\mathcal{X}}
\newcommand{\ibt}{t}
\newcommand{\ibT}{\myrv{T}}
\newcommand{\ibtalp}{\mathcal{T}}

\newcommand{\kappav}{\kappa^\labelvn}
\newcommand{\kappac}{\kappa^\labelcn}
\newcommand{\kappch}{\kappa^\labelch}

\newcommand{\quant}{Q}

\newcommand{\p}{p} %
\newcommand{\opllr}{L}
\newcommand{\neglogtanh}{\varphi}
\newcommand{\nch}{n^\labelch}
\newcommand{\rch}{r^\labelch}

\newcommand{\matH}{\mathbf{H}}
\newcommand{\matHb}{\matH_\mylabel{b}}
\newcommand{\Hb}{\mathsf{H}_\mylabel{b}}
\newcommand{\matHbr}{\matH_\mylabel{b}^{(\myr)}}
\newcommand{\idxi}{i}
\newcommand{\idx}{\mathsf{idx}}
\newcommand{\idxj}{j}
\newcommand{\iZz}{\idxj\Z{+}\z}
\newcommand{\setJ}{\mathcal{J}}
\newcommand{\Hbij}{\Hb^{\idxi\idxj}}
\newcommand{\Mb}{M_\mylabel{b}}
\newcommand{\J}{J}
\newcommand{\opcol}{\operatorname{col}}
\newcommand{\oprow}{\operatorname{row}}
\newcommand{\opsgn}{\operatorname{sgn}}
\newcommand{\B}{\mathsf{B}}
\newcommand{\vecb}{\myvec{b}}

\newcommand{\myr}{r}
\newcommand{\dc}{\mathsf{d}^\labelcn}
\newcommand{\dv}{\mathsf{d}^\labelvn}
\newcommand{\setN}{\mathcal{N}}
\newcommand{\n}{n}
\newcommand{\opNv}{\operatorname{N}^\labelvn}
\newcommand{\opNc}{\operatorname{N}^\labelcn}
\newcommand{\tilden}{\tilde{n}}
\newcommand{\alphastar}{\alpha_\star}

\newcommand{\setAc}{\mathcal{A}^\labelcn}
\newcommand{\setAstar}{\mathcal{A}^\star}
\newcommand{\alphav}{\alpha_\labelvn} %
\newcommand{\alphac}{\alpha_\labelcn}
\newcommand{\setU}{\mathcal{U}}
\newcommand{\setUv}{\mathcal{U}^\labelvn}
\newcommand{\setUc}{\mathcal{U}^\labelcn}
\newcommand{\Uinit}{\mytuple{\myset{U}}_\mylabel{init}}
\newcommand{\tuplesetU}{\mytuple{\myset{U}}}
\newcommand{\updateidx}{k}
\newcommand{\itermax}{\iota_\mylabel{max}}
\newcommand{\iteravg}{\iota_\mylabel{avg}}
\newcommand{\lch}{\ell^\labelch}
\newcommand{\lchmax}{\lch_\mylabel{max}}
\newcommand{\Lch}{\myrv{L}^\labelch}
\newcommand{\tch}{t^\labelch}
\newcommand{\Tch}{\myrv{T}^\labelch}
\newcommand{\tchalp}{\mathcal{T}^\labelch}
\newcommand{\wch}{w^\labelch}
\newcommand{\Tc}{\myrv{T}^\labelcn}
\newcommand{\tc}{t^\labelcn}

\newcommand{\Tv}{\myrv{T}^\labelvn}
\newcommand{\tv}{t^\labelvn}
\newcommand{\tvalp}{\mathcal{T}^\labelvn}
\newcommand{\Tstar}{\myrv{T}^\star}
\newcommand{\tstar}{t^\star}
\newcommand{\lstar}{\ell^\star}

\newcommand{\tstaralp}{\mathcal{T}^\star}

\newcommand{\ystaralp}{\mathcal{Y}^\star}
\newcommand{\barTv}{\bar{\myrv{T}}^\labelvn}

\newcommand{\barTc}{\bar{\myrv{T}}^\labelcn}

\newcommand{\yc}{\myvec{y}^\labelcn}
\newcommand{\Yc}{\myvec{\myrv{Y}}^\labelcn}
\newcommand{\ycalp}{\mathcal{Y}^\labelcn}
\newcommand{\yv}{\myvec{y}^\labelvn}
\newcommand{\Yv}{\myvec{\myrv{Y}}^\labelvn}
\newcommand{\barYstar}{\myvec{\bar{\myrv{Y}}}^\star}

\newcommand{\barystar}{\myvec{\bar{y}}^\star}
\newcommand{\barTstar}{\bar{\myrv{t}}^\star}

\newcommand{\barXstar}{\bar{\myrv{x}}^\star}

\newcommand{\barxstar}{\bar{x}^\star}

\newcommand{\yvalp}{\mathcal{Y}^\labelvn}

\newcommand{\x}{x} %
\newcommand{\X}{\myrv{X}}

\newcommand{\barX}{\bar{\X}}

\newcommand{\Eb}{E_b}
\newcommand{\No}{N_0}
\newcommand{\Es}{E_s}
\newcommand{\EbNo}{\Eb/\No}
\newcommand{\I}{\operatorname{I}}
\newcommand{\DeltaIv}{\Delta\!\I^\labelvn}
\newcommand{\DeltaIc}{\Delta\!\I^\labelcn}
\newcommand{\lv}{\ell^\labelvn}
\newcommand{\Lv}{\myrv{L}^\labelvn}
\newcommand{\lc}{\ell^\labelcn}

\newcommand{\Lc}{\myrv{L}^\labelcn}
\newcommand{\uLc}{\underline{\myrv{L}}^\labelcn}
\newcommand{\uLv}{\underline{\myrv{L}}^\labelvn}

\newcommand{\barLc}{\bar{\myrv{L}}^\labelcn}

\newcommand{\barLv}{\bar{\myrv{L}}^\labelvn}

\newcommand{\oprnd}{\operatorname{rnd}}
\newcommand{\phic}{\phi^\labelcn}
\newcommand{\phiv}{\phi^\labelvn}
\newcommand{\phich}{\phi^\labelch}
\newcommand{\phiminus}{\phi^{-}}
\newcommand{\tcminus}{t^{\labelcn,-}}
\newcommand{\VNplus}{\operatorname{VN}^+}
\newcommand{\VNminus}{\operatorname{VN}^-}

\newcommand{\TPblk}{\operatorname{TP}_{\mathsf{blk}}}
\newcommand{\TProw}{\operatorname{TP}_{\mathsf{row}}}
\newcommand{\TP}{\operatorname{TP}}
\newcommand{\Nd}{N_\mathsf{d}}
\newcommand{\setNr}{\mathcal{N}^{(r)}}
\newcommand{\Ndblk}{N_\mathsf{d}^\mathsf{blk}}
\newcommand{\Ndrow}{N_\mathsf{d}^\mathsf{row}}
\newcommand{\Nlayersr}{N_{\mathsf{layers}}^{(r)}}
\newcommand{\mone}{\mathsf{m}_1}
\newcommand{\idxone}{\mathsf{idx}_1}
\newcommand{\mtwo}{\mathsf{m}_2}
\newcommand{\mmag}{\mathsf{m}}
\newcommand{\Ag}{A_\mathsf{g}}

\newcommand{\dcR}{d^\mathsf{R}_\labelcn}
\newcommand{\lmax}{\ell_\mathsf{max}}
\newcommand{\AEmy}{\operatorname{AE}}
\newcommand{\Zp}{Z_\mylabel{p}}
\newcommand{\dummyvar}{\zeta}
\newcommand{\bmtau}{\mytuple{\tau}}
\newcommand{\bmphi}{\mytuple{\phi}}

\definecolor{mylightgreen}{RGB}{87, 189, 195}
\definecolor{myblue}{RGB}{0, 102, 204}
\definecolor{myred}{RGB}{204, 51, 51}
\definecolor{mylightred}{RGB}{255, 128, 128}
\definecolor{mygreen}{RGB}{51, 153, 51}
\definecolor{mylightgreen2}{RGB}{155, 187, 89}
\definecolor{myorange}{RGB}{236, 110, 0}

\newacronym[description={Additive White Gaussian Noise}]{AWGN}{AWGN}{additive white Gaussian noise}
\newacronym[description={Log-Likelihood Ratio}]{LLR}{LLR}{log-likelihood ratio}
\newacronym[description={Check Node}]{CN}{CN}{check node}
\newacronym[description={Variable Node}]{VN}{VN}{variable node}
\newacronym[description={VN-to-CN}]{VTC}{VTC}{VN-to-CN}
\newacronym[description={CN-to-VN}]{CTV}{CTV}{CN-to-VN}
\newacronym[description={Inter-Symbol Interference Node}]{ISI}{ISI}{inter-symbol interference}
\newacronym[description={BPSK}]{BPSK}{BPSK}{binary-phase shift-keying}
\newacronym[description={BP}]{BP}{BP}{belief propagation}
\newacronym[description={IB}]{IB}{IB}{information bottleneck}
\newacronym[description={APP}]{APP}{APP}{a-posteriori probability}
\newacronym[description={LUT}]{LUT}{LUT}{lookup table}
\newacronym[description={NMS}]{NMS}{NMS}{normalized min-sum}
\newacronym[description={OMS}]{OMS}{OMS}{offset min-sum}
\newacronym[description={uVTC}]{UVTC}{uVTC}{unquantized VTC}
\newacronym[description={B-APP}]{BAPP}{B-APP}{block-parallel APP}
\newacronym[description={R-APP}]{RAPP}{R-APP}{row-parallel APP}
\newacronym[description={R-VC}]{RVC}{R-VC}{row-parallel VC}
\newacronym[description={FER}]{FER}{FER}{frame error rate}
\newacronym[description={LDPC}]{LDPC}{LDPC}{low-density parity-check}

\def\mymodestandalone{buildmissing}

\usepackage{hyperref}
\usepackage[capitalise]{cleveref}
\crefformat{equation}{(#2#1#3)}
\crefrangeformat{equation}{(#3#1#4--#5#2#6)}
\crefmultiformat{equation}{(#2#1#3)}{ and~(#2#1#3)}{, (#2#1#3)}{, and~(#2#1#3)}

\crefname{figure}{Fig.}{Figs.} %
\Crefname{figure}{Fig.}{Figs.} %
\crefname{algorithm}{Alg.}{Algs.} %
\Crefname{algorithm}{Alg.}{Algs.} %

\crefname{line}{Line}{Lines} %
\Crefname{line}{Line}{Lines} %
\crefrangeformat{line}{Lines #3#1--#2#4}

\newcommand{\linelabel}[1]{\phantomsection\label{#1}}
\begin{document}

\title{%
Region-Specific Coarse Quantization With\\Check Node Awareness in 5G LDPC Decoding
\vspace{-0.2cm}
}
\author{
	Philipp Mohr\textsuperscript{\orcidlink{0000-0003-4350-9969}},
	\IEEEmembership{Graduate Student Member, IEEE}, and Gerhard Bauch, \IEEEmembership{Fellow, IEEE}
	\thanks{Received 19 June 2024; revised 16 November 2024 and 21 January 2025; accepted 28 January 2025. The associate editor coordinating the review of this article and approving it for publication was E. Dupraz. (Corresponding author: Philipp Mohr.)
		
The authors are with the Institute of Communications, Hamburg University of Technology, Hamburg, 21073, Germany. e-mail: \{philipp.mohr; bauch\}@tuhh.de.
	
	Digital Object Identifier 10.1109/TCOMM.2025.3541036}
}

\markboth{IEEE TRANSACTIONS ON COMMUNICATIONS}%
{How to Use the IEEEtran \LaTeX \ Templates}
\maketitle

\begin{abstract}
This paper presents novel techniques for improving the error correction performance and reducing the complexity of coarsely quantized 5G-LDPC decoders.
The proposed decoder design supports arbitrary message-passing schedules on a base-matrix level by modeling exchanged messages with entry-specific discrete random variables.
Variable nodes (VNs) and check nodes (CNs) involve compression operations designed using the information bottleneck method to maximize preserved mutual information between code bits and quantized messages.
We introduce alignment regions that assign the messages to groups with aligned reliability levels to decrease the number of individual design parameters.
Group compositions with degree-specific separation of messages improve performance by up to 0.4\,dB. Further, we generalize our recently proposed CN-aware quantizer design to irregular LDPC codes and layered schedules.
The method optimizes the VN quantizer to maximize preserved mutual information at the output of the subsequent CN update, enhancing performance by up to 0.2\,dB.
A schedule optimization modifies the order of layer updates, reducing the average iteration count by up to 35\,\%.
We integrate all new techniques in a rate-compatible decoder design by extending the alignment regions along a rate-dimension.
Our complexity analysis shows that 2-bit decoding can double the area efficiency over 4-bit decoding at comparable performance.
\end{abstract}

\vspace{-0.3cm}
\section{Introduction}\label{sec:introduction}
\newcommand\myfirstpagebottomspacing{2.0}
\enlargethispage{-\myfirstpagebottomspacing\baselineskip} %
\IEEEPARstart{S}{ince}  their invention by Gallager~\cite{gal62} and rediscovery by MacKay~\cite{mackay1997near}, \gls{LDPC} codes have become a crucial component in the field of modern communication systems. 
One important application is 5G New Radio (NR)~\cite{3gpp18}, where LDPC codes can achieve near-capacity performance with belief propagation decoding~\cite{gal62,mackay1997near}.

However, a significant bottleneck in belief propagation is the extensive exchange of messages between \glspl{VN} and \glspl{CN}.
Therefore, many works focused on reducing the bit width of the exchanged messages through quantization~\cite{thorpe03,lee05,kur08,kur14,romero16,lew18,meidlinger20,stark20,he19,mohr21,mohr22uniform, mohr22aware,mohr22low,monsees22,kang22,wang22,kumar13,ly17,lyu23,chen02,chen05,geiselhart_learning_2022}.
Conventional techniques approximate the belief propagation algorithms with sub-optimal quantization operations aiming at low-complexity implementations~\cite{chen05}.
A different approach is to design operations that maximize preservation of mutual information within the exchanged messages~\cite{thorpe03,lee05,kur08,kur14,romero16, lew18,meidlinger20,stark20,he19,mohr21,mohr22uniform, mohr22aware,mohr22low,monsees22,kang22,wang22}.
These decoders can achieve excellent performance at lower message resolutions than the conventional techniques.

Quantized belief propagation was introduced by Thorpe~\cite{thorpe03} where node operations aim at maximizing preserved mutual information between code bits and exchanged messages. Each node operation maps incoming messages onto a computational domain, combines them into a scalar via arithmetic operations, and then applies quantization, resulting in a compressed output message~\cite{lee05}.
Kurkoski \textit{et al.}~\cite{kur08} advanced this idea by replacing the mapping, arithmetic, and quantization steps with a cascade of two-input \glspl{LUT}, enabling compression inside a node.

\enlargethispage{-\myfirstpagebottomspacing\baselineskip} 

For the computational domain and the \gls{LUT} decoders the fundamental challenge is to design compression mappings that assign observed messages to a compressed message under preservation of relevant information.
The \gls{IB} method describes this problem with a relevant, observed, and compressed variable and can design compression mappings that maximize the mutual information between the compressed variable and the relevant variable~\cite{tishby00,lew18, kur14}. 

A further challenge arises when a decoder stage includes nodes with different degrees, resulting in messages with different alphabets of reliability levels.
Addressing these alphabets across subsequent decoder stages can lead to an unmanageable variety of \glspl{LUT}.
\emph{Message alignment} was proposed in~\cite{lew17mes} which can be used to design decoder stages so that output messages refer to the same reliability alphabet.
The technique enabled rate-compatible LUT decoders for 5G codes in~\cite{stark20} with aligned messages between consecutive decoder stages.

He \textit{et~al.}~\cite{he19} combined the original quantized belief propagation approach from~\cite{lee05} with the information-optimum algorithm from~\cite{kur14}. 
The combination resulted in better performance than the two-input LUT decoders which suffer from consecutive compression steps within each node update.

Another highly relevant aspect is the composition and order of decoder stages determined by the decoding schedule.
A layered schedule subdivides the parity-check matrix into layers and can halve the required number of message updates compared to a flooding schedule~\cite{hocevar04}.
Our works~\cite{mohr21,mohr22low} extended the mutual information maximizing decoding algorithms to row- and column-layered decoding for regular LDPC codes with computational domain and LUT decoders.
Kang \textit{et~al.}~\cite{kang22} developed a computational domain decoder design that supports column-layered decoding of irregular LDPC codes.
Lv \textit{et~al.}~\cite{lv22} proposed an entry-specific decoder design but investigated only a high-rate code.

This work develops mutual information maximizing techniques focusing on 5G LDPC codes~\cite{3gpp18}.
The wide range of node degrees and required support for rate compatibility make the decoder design for very coarse quantization challenging.
The following contributions aim to improve the current state-of-the-art such that performance loss resulting from coarse quantization is minimized.

\cref{sec:enc_dec_5g,sec:dde_model,sec:message_alignment_regions} contribute a design framework for 5G LDPC decoding. The framework makes use of a probabilistic memory model that characterizes the exchanged messages on a base-matrix level with probability mass functions.
The model is able to track the evolution of probability mass functions under arbitrary decoding schedules. %
One key difference compared to other works is the introduction of \emph{alignment regions} in \cref{sec:message_alignment_regions}.
Those regions define subsets of exchanged messages which share the same alphabet of reliability levels.
Thus, our framework supports configurations with different sets of reliability levels within a single update step of the decoding schedule.	
The quantization functions are designed to maximize mutual information between bit and message variables defined through the alignment regions.

We note that other works have strong similarities to our framework in certain configurations. For example, \cite{lv22} implicitly designs quantization of \gls{VN} messages with row-alignment and reconstruction of \gls{CN} messages with entry-specific alignment under a row-layered schedule. 
Furthermore, \cite{kang22} implicitly designs \gls{VN} and \gls{CN} updates both with matrix-alignment under a flooding schedule or with column-alignment under a column-layered schedule.
In addition,~\cite{wang22} implicitly designs \gls{VN} and \gls{CN} updates with row-alignment under a row-layered schedule.
Our work unifies all these decoders in a single framework.
Further, entirely new configurations lead to substantially improved performance.
For example, a flooding schedule decoder is significantly enhanced by designing degree-specific quantizers within one iteration using column- and row-alignment regions.
A compromise between matrix and degree-specific alignment is proposed as the \mbox{matrix-2} alignment using only two regions.
Messages from degree-one \glspl{VN} do not improve their reliability during decoding.
Thus, the first region excludes rows connected to degree-one \glspl{VN}, for allowing a wider range of reliability levels across iterations.

A further key contribution, introduced in \cref{sec:cnaware_quantization}, is an overall improved quantizer design method which is able to outperform the mutual information maximizing design from~\cite{kang22,lv22}, and \cite{wang22}.
The design method considers the subsequent processing of quantized \gls{VN} messages in a min-sum \gls{CN} update.
The fundamental idea for this approach was presented in our previous work~\cite{mohr22aware} but with the restriction to regular LDPC codes with flooding schedules and uniform quantization. 
The reformulation of this work allows the application of the \gls{CN}-aware design method to irregular LDPC codes decoded with arbitrary schedules.
Furthermore, our reformulation enables application of the \gls{IB} method to design non-uniform quantization thresholds.
Finally, we provide an information-theoretic analysis which explains the different optimization results compared to the \gls{CN}-\emph{unaware} optimization as in~\cite{kang22},\cite{lv22}, and \cite{wang22}.
Simulation results confirm excellent performance improvements particularly for very low-resolution decoding.

Another aspect that has not been considered in the closely related literature is the optimization of a layered decoding scheme. 
In \cref{sec:layeredopt}, we extend the developed design framework with an optimization step that selects the layer to be updated next that yields the highest mutual information gain.
The average number of messages to be exchanged is reduced by up to 35\,\% for row-layered decoding.
We note that other works, e.g. \cite{wang_two_2020}, provide solutions for optimized static and dynamic decoding schedules.
However, to the best of our knowledge, our work is the first to propose a static schedule embedded in a coarsely quantized decoder design.

As 5G makes use of incremental redundancy transmission and must work under a wide range of channel SNRs, a universal decoder is required to handle varying code rates. 
We show that our proposed techniques can be integrated in a rate-compatible decoder with very small performance loss over a wide range of code rates by expanding the alignment regions along a rate-dimension. 
The approach averages rate-specific probabilities of the exchanged messages for the design of common quantizers and reconstructions similar to~\cite{kang_design_2022} and~\cite{kang_memory_2022}.

Finally, we investigate the space and time complexity in \cref{sec:complexity} with a special focus on low resolutions and how hardware implementation can benefit from them.
Therefore, we evaluate the complexity based on state-of-the-art hardware architectures\cite{lee22multi},\cite{jang24area},\cite{ren24}.
The analysis shows that 2-bit decoding can double the area-efficiency compared to 4-bit decoding.

\vspace{-0.1cm}
\section{Encoding and Decoding of 5G LDPC Codes}\label{sec:enc_dec_5g}
\begin{figure}[t]
	\centering
	\subfloat[\centering Puncturing parity bits disables parts of a rate-compatible template matrix, leading to a rate-specific base matrix $\matHb$.]{
\includestandalone[mode=buildmissing]{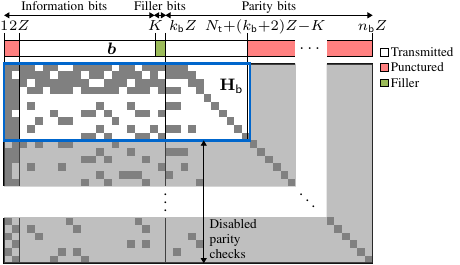}
		\label{fig:basegraph1_encoding}}
	\hfil
	\subfloat[\centering Matrix $\matH$ after lifting $\matHb$ with $Z{=}20$.]{
		\resizebox{0.99\linewidth}{!}{\includegraphics[width=1.0\linewidth]{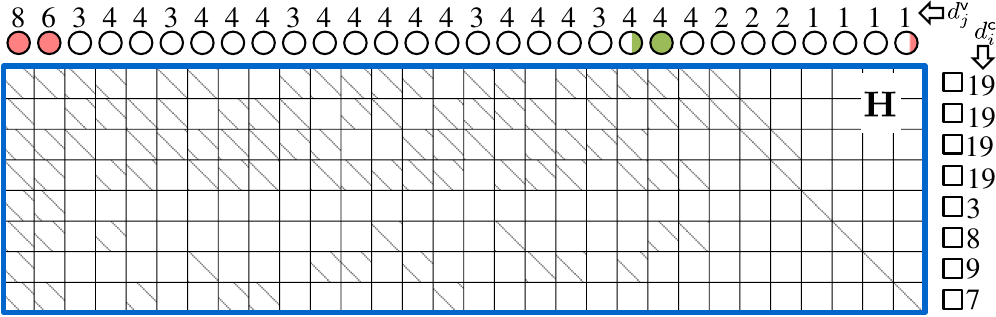}}\label{fig:basegraph1lifted}
		}
	\hfil
	\subfloat[\centering Tanner graph where the circles are \glspl{VN} and the squares are \glspl{CN}.]{
		\resizebox{0.99\linewidth}{!}{\includegraphics[width=1\linewidth]{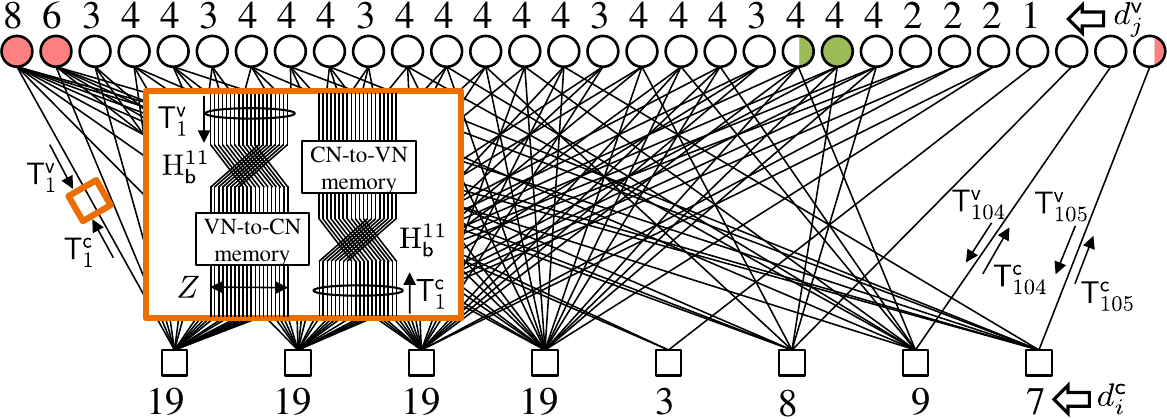}}
		\label{fig:tannergraph}}
\caption{Representations of 5G LDPC codes.}
\label{fig:representations_5G_ldpc_codes}
\end{figure}
\begin{figure*}[ht]
	\centering
	\includegraphics[width=0.9\linewidth]{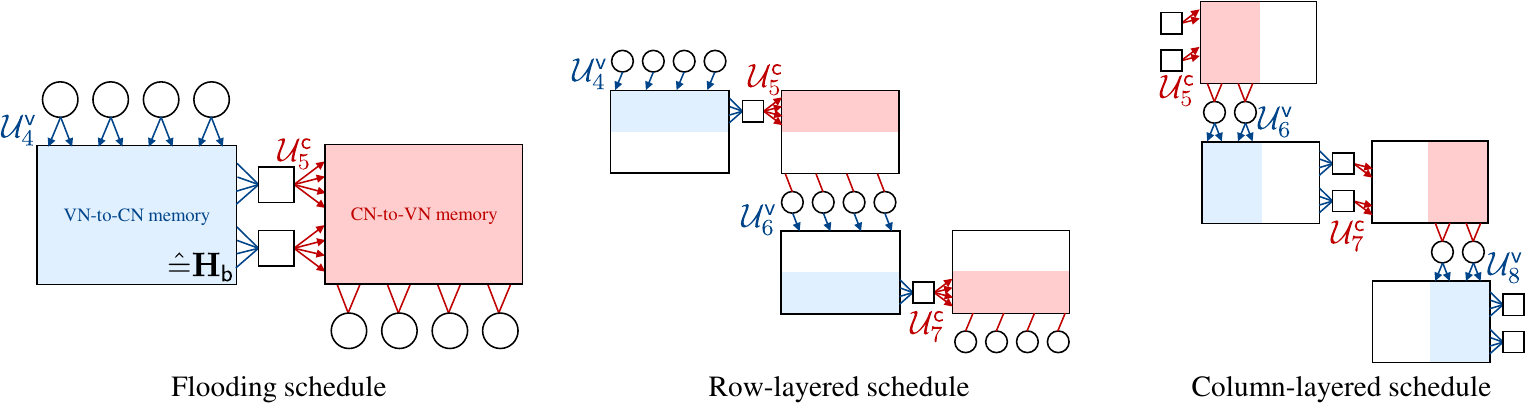}
	\vspace{-0.2cm}
	\caption{The flooding schedule updates all \gls{VN} memory locations~$\setUv_4$ followed by updating all \gls{CN} memory locations~$\setUc_5$. The row-layered schedule updates the memory locations corresponding to the first layer $\setUv_4$ and $\setUc_5$. Updates in the second layer $\setUv_6$ and $\setUc_7$ exploit improved reliability information from the first layer.
		In the column-layered schedule \gls{CN} updates $\setUc_5$ are followed by \gls{VN} updates $\setUv_6$, which is most efficient if all \gls{VN} memory locations are initialized. }
\label{fig:exampleschedules}
\end{figure*}
This section describes matrix and graph representations of 5G LDPC codes.
We use these representations to define LDPC decoding with arbitrary message passing schedules, such as a flooding or layered schedule.
Further, an efficient initialization schedule is introduced to avoid useless decoder operations due to punctured code bits.

Most standards define LDPC codes through a base matrix ${\matHb\in \{-1,\ldots,\Z{-}1\}^{\Mb\times \J}}$ highlighted in \cref{fig:basegraph1_encoding} for a 5G LDPC code.
The row and column indices are $\idxi{\in}\{1,\ldots,\Mb\}$ and ${\idxj{\in}\setJ=\{1,\ldots,J\}}$, respectively.
A lifting procedure replaces every entry $\Hbij$ of $\matHb$ with a $\Z{\times} \Z$ square matrix to obtain the full parity-check matrix $\matH$ as shown in \cref{fig:basegraph1lifted}.
The non-negative entries $\Hbij$ (gray, \cref{fig:basegraph1_encoding}) turn into cyclic $\Hbij$-right-shifted identity matrices.
The other entries (white) turn into zero matrices.
The encoder maps the information bits $\myvec{u}{\in}\{0, 1\}^{\myK}$ to code bits $\vecb=[\breal_{1,1},\ldots,\breal_{\J,\Z}]^\mytranspose{\in}\{0, 1\}^{\J\Z}$ such that $\matH\myvec{b}=\myvec{0}$.

In 5G NR the base matrix $\matHb$ is obtained from one of the two template base matrices (termed base graph~1~and~2) depending on the desired code rate~$r$ as well as the desired information block length~$\myK$~\cite{3gpp18}.
Base graph~2 is used if $\myK{\le}292$, $\myr{\le}0.25$,  or $\myr{\le}0.67$ and $\myK{\le}3824$.
Otherwise, base graph 1 is selected.
\cref{fig:basegraph1_encoding} depicts base graph~1 designed for rates $\myr{=}\nicefrac{1}{3}$ to $r{=}\nicefrac{22}{24}$~\cite{dahlman23}.
The maximum lifting size is chosen from a table satisfying $\myK{\leq} \kb \Z$ where $\kb{=}22$.
As indicated in \cref{fig:basegraph1_encoding}, the information bits are placed at positions~$1$~to~$\myK$.
Filler bits which are known by the receiver fill up the code word at positions $\myK{+}1$~to~$\kb \Z$ if $\myK<\kb \Z$.

The parity bits are computed using the information and filler bits~\cite{nguyen_efficient_2019}.
Only $\Nt{=}\lceil \myK/\myr\rceil$ bits of the full code word are selected for transmission (cf.\,\cref{fig:basegraph1_encoding}).
Parity checks with two punctured bits are disabled, reducing the size of the base matrix~$\matHb$ to ${\Mb=2+\lceil\nicefrac{(\Nt-\myK)}{\Z}\rceil}$ rows and ${\J=\kb+\Mb}$ columns.

For every $\breal_{\idxj,\z}$ the decoder observes a channel \gls{LLR} $\lch_{\idxj,\z}$.
Those LLRs are $0$ for the punctured bits, $\infty$ for the filler bits, and $\opllr(\breal|\rch)=\log\nicefrac{\p(\breal=0|\rch)}{\p(\breal=1|\rch)}$ for every transmitted bit $\breal$.
This work considers modulation to \gls{BPSK} symbols and \gls{AWGN} $\nch$ leading to $\rch{=}(2b{-}1){+}\nch$.

The base matrix ~$\matHb$ can be represented by a Tanner graph illustrated in \cref{fig:tannergraph}.
Each column~$\idxj$ turns into a \gls{VN} and each row~$\idxi$ into a \gls{CN}.
The edges between \glspl{VN} and \glspl{CN} represent non-negative entries $\Hbij$.
The node degree, i.e., the number of connected edges to a node, is $\dv_\idxj$ for a \gls{VN} and $\dc_\idxi$ for a \gls{CN}.

Message passing decoding computes and exchanges messages between \glspl{VN} and \glspl{CN} to aggregate extrinsic probability information from the parity-check constraints for error correction.
Each edge of the graph contains a \gls{VN} and \gls{CN} memory location enumerated with $\n{\in}\setN{=}\{1,\ldots,\sum_\idxj \dv_\idxj\}$ for \acrlong{VTC} and \acrlong{CTV} messages, respectively.
A memory location stores $\Z$ messages after lifting the graph as illustrated within the orange box in \cref{fig:tannergraph}.
The sets $\setUv{\subseteq}\setN$ and $\setUc{\subseteq}\setN$ specify target memory locations for \gls{VN} and \gls{CN} updates, respectively.
The decoding schedule defines the order in which memory locations are updated as
\begin{align}
	\tuplesetU=(\setUv_{0},\setUc_{1},\setUv_{2},\setUc_{3},\ldots)
\end{align}
followed by a final hard decision update that uses the most recently updated \gls{CN} messages.
\cref{fig:exampleschedules} illustrates three schedule types for one decoder iteration after initialization. 
One iteration involves updating a total of $2|\setN|$ memory locations.
Thus, the iteration count after each update step $k\in\{0,\ldots,|\tuplesetU|{-}1\}$ is
\begin{align}
	\iota(k)=\frac{1}{2|\setN|}\sum_{k'=0}^{k}{|\setU^\star_{k'}|} \text{ with $\star\in\{\labelvn,\labelcn\}$}.
	\label{equ:itercount}
\end{align}

\begin{algorithm}[t]
	\caption{Initialization Schedule for 5G LDPC Codes}\label{alg:initsched}
	\begin{algorithmic}[1]
		\footnotesize
		\Require memory locations $\setN$
		\Ensure $\Uinit=(\setUv_0,\setUc_1,\ldots)$
		\State $\setUv_0=\setUc_1=\{\n\in\setN:\opcol(\n)=1\}$
		\State $\setUv_2=\setUc_3=\{\n\in\setN:\opcol(\n)=2\}$
		\If{initialization for column-layered schedule}
		\State $\setUv_4=\setN-(\setUv_0\cup\setUv_2)$
		\EndIf
	\end{algorithmic}
\end{algorithm}
In the case of 5G LDPC codes the channel messages related to the first two columns are zero-LLRs due to puncturing~(cf. \cref{fig:representations_5G_ldpc_codes}).
Hence, updating the corresponding \glspl{VN} would create zero-LLR \gls{CN} inputs. 
A \gls{CN} update with at least a single extrinsic zero-LLR input leads to a zero-LLR output that is useless in further processing~\cite{ha07}. 
We carry out a fixed initialization sequence that omits \gls{CN} updates connected to punctured \glspl{VN}, avoiding useless updates.
A simple but effective initialization is proposed in \cref{alg:initsched}.
The initialization procedure saves the complexity of about 1.5 decoding iterations compared to standard schedule processing.

\section{Entry-Specific Quantized Decoder Design Using the Information Bottleneck Method}\label{sec:dde_model}

This section introduces a probabilistic design framework for a channel quantizer and the decoder operations with coarse quantization.
Any update of a \gls{VN} or \gls{CN} memory location~$\n$ results in $\Z$ updated messages as shown in \cref{fig:tannergraph}.
Thus, limiting the number of bits for each message is crucial for reducing the memory and routing complexity.
The operations are aware of the meaning of any $w$-bit message $\ibt{\in}\{-2^{w-1},\ldots,-1,1,\ldots,2^{w-1}\}$, i.e., the probability $p(\x|\ibt)$ w.r.t.\ a relevant code bit $\x{\in}\{0,1\}$.
The design goal for any operation is to maximize mutual information $\I(\ibX;\ibT)$.

As a first step, \cref{sec:mim_channel_quant} designs a channel quantizer.
The bits $\breal_{\idxj,\z}$ and channel \glspl{LLR} $\lch_{\idxj,\z}$ are modeled with the random variables $\B_\idxj$ and $\Lch_\idxj$ for every base-matrix column~$\idxj$.
The design is initialized using $\p(\breal_\idxj,\lch_\idxj)$ obtained with \cref{alg:chdist}.
The channel quantizer outputs are modeled with the random variables~$\Tch_\idxj$.

The exchanged messages between \glspl{VN} and \glspl{CN} in \cref{fig:tannergraph} are modeled using discrete random variables $\Tv_\n$ and $\Tc_\n$ for every memory location $\n$.
Each $\n$ can be associated with a relevant bit variable $\X_\n= \B_{\opcol(\n)}$.
A node update changes the meanings of updated messages, $\p(\x_\n|\tv_\n)$ or $\p(\x_\n|\tc_\n)$, which are tracked via discrete density evolution~\cite{chen02}.
Table~\ref{table:randomvariables} summarizes the properties of random variables used in the decoder model.
\cref{sec:vndesign,sec:cndesign} describe mutual information maximizing updates of any \gls{VN} or \gls{CN} memory location~$\n$.
The design takes into account \emph{location-dependent} meanings of messages resulting from the schedule, node degrees, punctured bits, and filler bits.
\begingroup %
\setlength{\tabcolsep}{4pt} %
\renewcommand{\arraystretch}{1.2} %
\begin{table}[t]
	\centering
	\caption{Random variables in the probabilistic model}
	\label{table:randomvariables}
	\vspace{-0.0cm}
	\includestandalone[mode=\mymodestandalone]{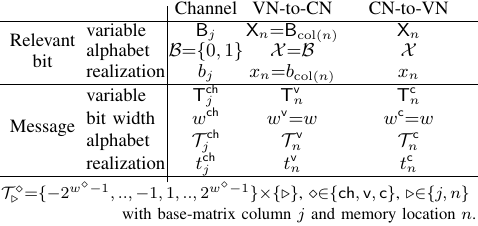}
\end{table}
\endgroup

\begin{algorithm}[t]
	\caption{Probability Distribution of Channel Messages}\label{alg:chdist}
	\begin{algorithmic}[1]
		\footnotesize
		\Require design-$\EbNo$, code rate $\myr$, fine resolution $\kappch$, maximum \gls{LLR} $\lchmax$
		\Ensure $\p(\breal_\idxj,\lch_\idxj)$
		\State For the punctured positions $\idxj{,}\z\in \setJ{\times}\setZ:\iZz{<}2\Z\text{ or }\allowbreak \iZz{>}\allowbreak \Nt{+}(\kb{+}2)\Z{-}\myK$ (cf.\,\cref{fig:basegraph1_encoding}) set $\p(\breal,\lch|\idxj,\z)=\operatorname{const.}$
		\State For the filler positions $\idxj{,}\z\in\setJ{\times}\setZ: \myK{<}\iZz{<}\kb\Z$ (cf.\,\cref{fig:basegraph1_encoding}) set $$\p(\breal,\lch|{\idxj,\z})=\begin{dcases} 1.0 & b=0,\lch=\lchmax\\0.0 & \operatorname{otherwise}
		\end{dcases}.$$
		\State For the remaining $\idxj{,}\z$ set ${p(\breal,\lch|\idxj{,}\z)=\p(\breal, \lch)}$ where $\lch=\allowbreak\tilde{Q}(\opllr(\breal|r^\labelch))$ for the output ${\rch=\operatorname{BPSK}(\breal){+}\nch}$  of a memoryless \acrshort{AWGN} channel with mapping $\operatorname{BPSK}(\breal)=2\breal{-}1$.
		The $\EbNo$ and code rate $r$ determine the variance of real-valued \acrshort{AWGN} $\nch$. %
		Quantization is applied with
		$\tilde{Q}(\ell)=\operatorname{sgn}(\ell)\max(\lfloor |\ell|/\kappch {+}\frac{1}{2}\rfloor,\lchmax)$ to obtain a probability mass function.
		\State Compute $p(\breal,\lch|j)=E_{z}\{p(\breal,\lch|\idxj{,}\z)\}$ by averaging over all $z$ that are input to base-column $j$.
	\end{algorithmic}
\end{algorithm}
\begin{algorithm}[t]
	\caption{Design of the Channel Quantizer}\label{alg:chdes}
	\begin{algorithmic}[1]
		\footnotesize
		\Require
		$\wch,\p(\breal_\idxj,\lch_\idxj)$ %
		\Ensure
		$\myvec{\tau}^\labelch$, $p(\breal_j,\tch_j)$
		\State Compute $p(\ibX=\ibxreal,\ibY=\ell)=\myopexpecation_{j}\{\p(\B_j=\ibxreal,\Lch_\idxj=\ell)\}$ $\forall x,\ell$.
		\State Design $\wch$-bit quantizer $Q$ maximizing $\I(\ibX;\quant(\ibY))$. %
		\State Compute
			$p(\breal_j,\tch_j)=\sum_{\lch_j}\delta(\tch_j{-}\quant(\lch_j))p(\breal_j, \lch_j)$ $\forall j \in \setJ$.
		\State Store thresholds $(\tau_{1},\ldots,\tau_{2^{w-1}{-}1})$ of $Q$ in $\myvec{\tau}^\labelch$.
	\end{algorithmic}
\end{algorithm}

\subsection{Mutual Information Maximizing Channel Quantization}\label{sec:mim_channel_quant}
For each bit $\breal_\idxj$, the channel quantizer observes a high-resolution LLR $\lch_\idxj$ that shall be quantized to a $\wch$-bit message $\tch_{j}$ with minimal loss of relevant information.
The messages $\breal_\idxj, \lch_\idxj \text{ and } \tch_{j}$ can be modeled by three discrete random variables $\ibX,\ibY \text{ and } \ibT$ that form a Markov chain $\ibX{\to}\ibY{\to}\ibT$.

This kind of setup is covered by the \gls{IB} framework where $\ibX, \ibx{\in} \ibxalp$ is the relevant variable, $\ibY, \iby{\in} \ibyalp$ is the observed variable, and $\ibT,\ibt{\in}\ibtalp$ is the compressed variable~\cite{lew18}.
From an information theoretic perspective, the mutual information $\I(\ibX;\ibY)$ upper bounds the achievable information rate. Thus, designing a compression mapping $\p(\ibt|\iby)$ with minimal mutual information loss $\I(\ibX;\ibY){-}\I(\ibX;\ibT)$ is considered as the information-optimum objective, i.e., $\max_{p(t|y)}\I(\ibX;\ibT)$.

In \cite{kur14} it was shown for a binary alphabet $\mathcal{X}{\in}\{0,1\}$ that the optimal deterministic mapping $p(t|y)$ can be defined using a set of quantization thresholds $\myvec{\tau}=(\tau_0,\ldots, \tau_{|\mathcal{T}|})$ according to
\begin{align}
	t=\quant(L(x|y))=
	\ibtalp[k] \quad%
	\tau_{k}{\leq}\opllr(x|y){<}\tau_{k+1}, 0{<}k{\le} |\mathcal{T}|
	\label{eq:mimquantizer}
\end{align}
with \gls{LLR} $\opllr(x|y)=\log(p(\ibX{=}0|y)/p(\ibX{=}1|y))$, outer thresholds $\tau_0={-}\infty$ and $\tau_{|\mathcal{T}|}={+}\infty$, and $\ibtalp[k]$ identifying the $k$-th element of the ordered set $\ibtalp$.

Only the joint distribution $p(x,y)$ and the compressed alphabet size $|\ibtalp|$ need to be specified before the optimization is performed.
As in ~\cite{mohr22uniform}, symmetric thresholds are enforced such that  $\tau_k{=}-\tau_{|\ibtalp|{-}k}$ and $\tau_{|\ibtalp|/2}{=}0$.
We design one quantization threshold set $\myvec{\tau}^\labelch$ for all columns $j$ according to \cref{alg:chdes}.

\begin{figure}[t]
	\centering
	\includestandalone[mode=\mymodestandalone]{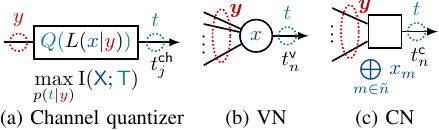} 
	\vspace{-0.1cm}
	\caption{The decoder operations channel quantization, \gls{VN} update and \gls{CN} update can be designed using the \gls{IB} method. The mapping $p(t|y)$ maximizes preserved mutual information between $\ibX$ and $\ibT$.}
	\vspace{-0.0cm}
	\label{fig:chvncn}
\end{figure}
\subsection{Entry-Specific Quantized Variable Node Design}\label{sec:vndesign}
In the probabilistic decoder model, \glspl{VN} update the joint probability mass functions $p(x_\n,\tv_\n)$ for all memory locations $n{\in}\setUv$ defined in a single step of the decoding schedule~$\mytuple{\myset{U}}$.
This section designs a specific \gls{VN} update for each $n{\in}\setUv$ to maximize mutual information~$\I(\ibX_n;\Tv_n)$.
A \gls{VN} update for a location $n$ in column ${j=\opcol(n)}$ with degree ${d=\dv_j}$ combines a channel message and extrinsic \gls{CN} messages, $\yv_\n=(\tch_{j},\tc_{\n_1},\ldots,\tc_{\n_{d-1}})$, into a compressed output message~$\tv_\n$.
The indices $\n_k$ refer to the extrinsic \gls{VN} locations in $\tilden=\opNv(j){\setminus}\{n\}$ where $\opNv(j)=\{n'{\in}\mathcal{N}{:}\opcol(n')=j\}$.
The observed inputs
\begin{align}
	(y_1,\ldots,y_d)=\yv_n\in\yvalp_n=\tchalp_j{\times}\tvalp_{\n_1}{\times} \ldots{\times} \tvalp_{\n_{d-1}}\label{eq:input_combination_vn}
\end{align}
are modeled by the random variable $\Yv_{\!\n}$ and provide extrinsic information about $\ibX_{n}$.
\cref{fig:chvncn}b shows a \gls{VN} IB setup, where $\ibX_\n$, $\Yv_{\!\n}$ and $\Tv_\n$ are the relevant, observed, and compressed variables $\ibX$, $\myvec{\ibY}$ and $\ibT$, respectively.
As in \cref{sec:mim_channel_quant}, an IB algorithm finds a mutual information maximizing compression mapping $p(t|\myvec{y})$ realized with threshold quantization $t=\quant(L(x|\myvec{y}))$ where~\cite{lee05}
\begin{align}
	L(x|\myvec{y})=L(x)+\sum\nolimits_{k=1}^{d}L(y_k|x).\label{eq:vn_computation}
\end{align}
We remark that \cref{eq:vn_computation} neglects cycles of the graph.
The hard decision for the code bit $\breal_j$ is ${\hatb_j=(1-\operatorname{sgn}(\hatl_j))/2}$ using the \gls{APP} \gls{LLR} $\hatl_{\opcol(n)}=L(x|\myvec{\iby}){+}\opllr(\tc_n|x)$.

In an implementation,~\eqref{eq:vn_computation} is considered as the computational part where the LLRs $L(y_k|x)$ are reconstructed using \glspl{LUT}.
The implementation and design complexity can be significantly reduced by processing $w'$-bit integer representations of the real-valued LLRs. 
The integer representations are obtained by scaling and rounding the underlying real numbers.

\subsection{Entry-Specific Quantized Check Node Design}\label{sec:cndesign}
In the probabilistic decoder model, \glspl{CN} update the joint probability mass function $p(\x_\n,\tc_n)$ for all memory locations $n{\in}\setUc$ defined in a single step of the decoding schedule~$\mytuple{\myset{U}}$.
This section designs a specific \gls{CN} update for each $n{\in}\setUc$ to maximize mutual information~$\I(\ibX_\n;\ibT^\labelcn_\n)$.
A \gls{CN} update for a location $n$ in row~${i=\operatorname{row}(n)}$ with degree ${d=\dc_i}$ combines
the extrinsic \gls{VN} messages $\yc_\n=(\tv_{\n_1},\ldots,\tv_{\n_{d-1}})$ into a compressed output message~$\tc_\n$.
The indices $\n_k$ point to the extrinsic \gls{VN} locations in $\tilden=\opNc(i){\setminus}\{n\}$ where $\opNc(i)=\{n'{\in}\mathcal{N}{:}\oprow(n')=i\}$.
The observed inputs
\begin{align}
	(y_1,\ldots, y_{d-1})=\yc_\n\in\ycalp_\n= \tvalp_{\n_1}\times \ldots\times \tvalp_{\n_{d-1}},\label{eq:input_combination_cn}
\end{align}
are modeled by the random variable $\Yc_{\!\n}$ and
provide extrinsic information about the variable $\ibX_\n$ constrained as $\X_\n=\X_{\n_{1}}\oplus\ldots\oplus \X_{\n_{d-1}}$.
\cref{fig:chvncn}c shows a \gls{CN} IB setup, where $\X_\n$, $\Yc_{\!\n}$ and $\Tc_\n$ are the relevant, observed and compressed variables $\ibX$, $\myvec{\ibY}$ and $\ibT$, respectively.
As in \cref{sec:mim_channel_quant}, an~IB algorithm finds a mutual information maximizing compression mapping $p(t|\myvec{y})$ realized with threshold quantization $t=\quant(L(x|\myvec{y}))$ where~\cite{lee05}
\begin{align} %
	\opllr(x|\myvec{y})&	%
	=\prod\nolimits_{k=1}^{d{-}1}
\operatorname{sgn}(L^\labelvn_{k})\cdot\neglogtanh^{-1}\left(\sum\nolimits_{k=1}^{d{-}1}\neglogtanh(|L^\labelvn_{k}|)\right)\label{eq:cn_computation} %
\end{align}
with the reconstructed inputs obtained through
\begin{align}
	L^\labelvn_{k}=\opllr(x_{\n_k}|y_k) \text{ and } \neglogtanh(|\ell|)={-}\log \tanh(|\ell|/2), \ell{\in} \mathbbR.\label{eq:neglogtanhfnc}
\end{align}
In an implementation, \cref{eq:cn_computation} is the computational part where the reconstruction uses a lookup table.
It holds that $\neglogtanh^{-1}(|\ell|)=\neglogtanh(|\ell|)$ which is monotonic decreasing w.r.t.~$|\ell|$.
Hence, computing $\neglogtanh^{-1}$ is not required as the threshold levels of~$Q$ can be adjusted to include the inverse transformation $\neglogtanh^{-1}$.

With the min-sum approximation, \cref{eq:cn_computation} simplifies to
\begin{align}
	L(x|\myvec{\iby})&\approx\prod\nolimits_{k=1}^{d{-}1}\operatorname{sgn}(L^\labelvn_{k})\min_{k}(|L^\labelvn_{k}|).\label{eq:cn_computation_min}
\end{align}
If all $y_k$ use a sign-magnitude format and have equal LLR levels, the reconstruction and quantization operation can be removed without affecting the behavior~\cite{kumar13,meidlinger20}:
\begin{align}
		t \approx \operatorname{MS}(\myvec{\iby})=\prod\nolimits_{k=1}^{d{-}1}\operatorname{sgn}(y_k)\min_{k}(|y_k|).\label{eq:cn_computation_min_approx}
\end{align}

\section{Node Design with Alignment Regions}\label{sec:message_alignment_regions}
\begin{figure}[t]
	\centering	
	\includestandalone[mode=buildmissing]{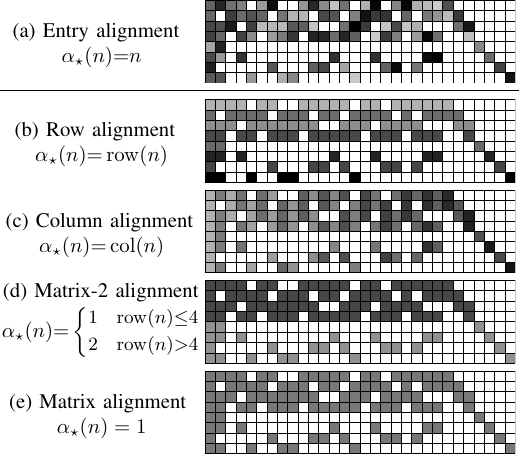}
	\vspace{-0.7cm}
	\caption{The alignment function $\alphastar$ organizes the base-matrix into distinct regions for \gls{VN} ($\star=\labelvn$) and \gls{CN} ($\star=\labelcn$) memory locations. Messages $t^\star_n$ associated with the same region $\alphastar(n)$ use the same set of reliability levels during reconstruction in the node operations.}
	\label{fig:alignment_regions}
\end{figure}

In \cref{sec:dde_model} we designed individual quantizers for each message variable $\Tstar_\n$ where ${\star=\labelvn}$ labels a \gls{VN} message and ${\star=\labelcn}$ labels a \gls{CN} message.
The message variables $\Tstar_\n$ model the exchanged decoder messages on a base-matrix level. The different grayscale tones in 
\cref{fig:alignment_regions}a illustrate potentially different reliability levels associated with $\Tstar_\n$ across the memory locations $n$.
The reliability levels $\p(\x_\n|\tstar_\n)$ of a message $\tstar_\n$ change with every node update and are mainly influenced by the node degrees and the decoding schedule.

This section introduces the mixture random variables $\barTstar_a$ and $\barXstar_a$, whose distributions are the average of the probability mass functions of all $\Tstar_n$ and $\X_n$ with $\alphastar(n) = a$:
\begin{align}
	p(\barXstar_a = x, \barTstar_a = t) 
	&= \myopexpecation_{n|a}\{p(\X_n = x, \Tstar_n = t)\}.
	\label{eq:avgvar}
\end{align}
This reduces the variety of decoder parameters for reconstruction and quantization, with $\barTstar_{\alphastar(n)}$ taking the role of~$\Tstar_n$ in the \emph{reconstructions}.
Specifically,
\begin{itemize} 
	\item in \eqref{eq:cn_computation} for VN messages ${\star=\labelvn}$  (see \cref{line:vnreconstruction} of \cref{alg:cndesigncomp}),
	\item in \eqref{eq:vn_computation} for CN messages $\star=\labelcn$   (see \cref{line:cnreconstruction} of \cref{alg:vndesign}).
\end{itemize}
The function ${\alphastar:\mathcal{N}{\to }\setAstar}$ assigns memory locations $n$ to regions enumerated with ${a{\in}\setAstar=\{1,\ldots,|\setAstar|\}}$.
Several averaging strategies are provided in \cref{fig:alignment_regions}b to~\ref{fig:alignment_regions}e where the same grayscale tone indicates membership to the same averaging region.

\subsection{Aligned Design of Memory Location Updates}
The averaging procedure can cause a performance loss as the averaged meaning $\p(\barXstar_{\alphastar(n)}{=}x|\barTstar_{\alphastar(n)}{=}t)$ for a message~$t$ w.r.t.\ bit~$\x$ can differ from its accurate meaning $\p(\X_n{=}x|\Tstar_n{=}t)$.
A joint design aligns the characteristics of variables $\Tstar_n$ within the averaging region~$\alphastar(n)$.
The underlying idea is termed \emph{message alignment} in literature~\cite{lew17mes,stark20}.
Hence, the averaging regions $\setAstar$ can also be seen as \emph{alignment regions}.
All extrinsic node input combinations are collected in the set $\ystaralp_n$  as defined in \cref{eq:input_combination_vn} for~$\star=\labelvn$ or \cref{eq:input_combination_cn} for~$\star=\labelcn$.
Any input combination in the alignment region $a$ is given by
\begin{align}
	\barystar_a\in\bar{\mathcal{Y}}^\star_a=\bigcup_{\n:\alphastar(n)=a} \ystaralp_{n}.
	\label{eq:alignment_observations}
\end{align}
The objective is to find a mapping $\p(\bar{t}^\star_a|\barystar_a)$ that maximizes $\I(\barXstar_a; \bar{\myrv{T}}^\star_a)$, solved using an IB algorithm as in \cref{sec:mim_channel_quant}.
The optimal mapping can be carried out with threshold quantization $\bar{t}^\star_a=Q(\opllr(\barxstar_a|\barystar_a))$.
Thus, the same quantizer thresholds can be used across the alignment region. 
The approach minimizes a mutual information loss $\I(\barXstar_{a};\barYstar_{\! a}){-}\I(\barXstar_{a};\bar{\myrv{T}}^\star_{a})$ resulting from quantization \emph{and} restriction to a single alphabet of reliability levels for all messages in region $a$.

\subsection{Modification for Layered Decoding}
Each step of a \emph{layered} decoding schedule in \cref{alg:decdesign} updates a subset of all memory locations $\setU^\star{\subset}\mathcal{N}$.
A quantizer design must only be performed for the regions with index $a{\in}\{\alphastar(n){:}n{\in}\setU^\star\}$ (cf. \cref{line:loopvnquantdesign} of \cref{alg:vndesign} and \cref{line:loopcnquantdesign} of \cref{alg:cndesigncomp}).
Not all locations in the function's preimage $\alphastar^{-1}(a)$ are necessarily updated with $\setU^\star$.
Thus, the quantizer designed for region $a$ is used only for updating the locations in $\setU^\star$ (cf. \cref{line:updatedistvtc} of \cref{alg:vndesign}).
\subsection{Alignment Strategies and Their Complexity}
\begin{figure}[t]
	\includestandalone[mode=\mymodestandalone]{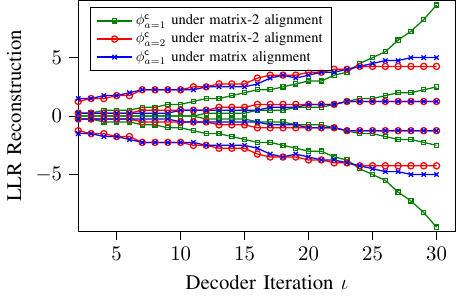}
	\vspace{-0.1cm}
	\caption{\gls{LLR} reconstruction levels $\kappav{\cdot}\phic_a(\tc)$ of 2-bit messages using matrix and matrix-2 alignment (flooding schedule, $r=\nicefrac{1}{3}$, $K=8448$, $\kappav=0.25$).}
	\vspace{-0.0cm}
	\label{fig:reconstructionlevels}
\end{figure}
\cref{fig:alignment_regions}b to \cref{fig:alignment_regions}e illustrate several alignment strategies.
\gls{VN} and \gls{CN} messages can use different strategies.
The row alignment in \cref{fig:alignment_regions}b preserves degree-specific reliability levels for the \gls{CN} messages.
Low-degree \glspl{CN} provide more reliable messages than high-degree \glspl{CN}.
Vice versa, the column alignment in \cref{fig:alignment_regions}c preserves degree-specific reliability levels for the \gls{VN} messages.
High-degree \glspl{VN} provide more reliable messages than low-degree \glspl{VN}.
The matrix alignment in \cref{fig:alignment_regions}e enforces the same reliability levels for all exchanged VN or \gls{CN} messages.
The matrix-2 alignment in \cref{fig:alignment_regions}d defines two complementary regions where the first region excludes rows connected to degree-one \glspl{VN}.
It is a compromise between the row and matrix alignment.
\cref{fig:reconstructionlevels} depicts the reconstruction functions $\phic_a$ (see \cref{line:cnreconstruction} of \cref{alg:vndesign}).
We observe that the reliability levels are much smaller under matrix-2 alignment for region ${a=1}$ in early iterations than for the matrix alignment.

The number of quantizers to be optimized per update under a flooding schedule for the matrix in \cref{fig:alignment_regions} is with entry alignment ${|\setAstar|=102}$, row alignment ${|\setAstar|=6}$, column alignment ${|\setAstar|=30}$, matrix-2 alignment $|\setAstar|=2$ and matrix alignment ${|\setAstar|=1}$.
Thus, the design complexity in \crefrange{line:loopvnquantdesign}{line:endloopvnquantdesign} of \cref{alg:vndesign} and \crefrange{line:loopcnquantdesign}{line:endloopcnquantdesign} of \cref{alg:cndesigncomp} reduces significantly.

For hardware implementations, the alignment determines the number of parameters that must be loaded for updating \glspl{VN} or \glspl{CN}.
Consider an example with a flooding schedule, an internal bit width $w'{=}8$\,bits and message bit width $w{=}3$\,bits. Each update uses reconfigurable LUTs to adjust parameters in the reconstruction and quantization operation.
The LUTs for reconstruction and quantizer thresholds are reconfigured with $(w'{-}1)2^{w-{1}}=24$\,bits and approximately $(w'{-}1)(w{-}1)=12$\,bits, respectively.
Then, one update loads $36|\setAstar|$\,bits for parameter reconfiguration.
In comparison, the number of message bits exchanged between \glspl{VN} and \glspl{CN} amounts to $2w|\setN|\Z=612\Z$\,bits.
For the investigated 5G \gls{LDPC} codes with $\Z=384$ our analysis suggests that the overhead from loading parameters is comparatively small with regard to the message transfers. 
However, as we will see in the next section, the alignment significantly influences the error correction performance under low-resolution decoding.

\subsection{Performance of Different Alignment Strategies}\label{subsec:fer_alignment}

\begin{algorithm}[t]
	\caption{Design of Decoder With Fixed Schedule}\label{alg:decdesign}
	\begin{algorithmic}[1]
		\footnotesize
		\Require $w, \tuplesetU,\alphac,\alphav, \p(\breal_j,\tch_j)$
		\Ensure $(\mytuple{\phi}_{a})_k$, $(\bmtau_{a})_k$
		\State We set the initial distributions $\p(\x_\n, \tstar_n)=(2|\tstaralp|)^{-1}$. The probability mass functions $\p(\x_\n, \tstar_n)$ change with every update step $k$.		
		\For{$\updateidx \in \{0,\ldots, |\tuplesetU|-1\}$}
		\If{$k$ is \gls{VN} update}
		\State $\mytuple{\phi}_{a},\mytuple{\tau}_{a}, p(\x_\n, \tv_\n) \gets$  Design update $\setUv_k$ with \cref{alg:vndesign} or \cref{alg:vndesignaware}.
		\ElsIf{$k$ is \gls{CN} update}
		\State $\mytuple{\phi}_{a},\mytuple{\tau}_{a},p(\x_\n, \tc_\n) \gets$  Design update $\setUc_k$ with \cref{alg:cndesigncomp} or \cref{alg:cndesignminsum}.
		\EndIf
		\EndFor
	\end{algorithmic}
\end{algorithm}

\begingroup
\setlength\belowdisplayskip{5pt} %
\setlength\abovedisplayskip{5pt} %
\begin{algorithm}[t]
	\caption{Design of Computational Domain \gls{VN} Update}\label{alg:vndesign}
	\begin{algorithmic}[1]
		\footnotesize
		\Require $w, \setUv,\alphac,\alphav,\p(\breal_j,\tch_j), \p(\x_\n, \tc_\n), \p(\x_\n, \tv_n), \kappav$
		\Ensure $\bmphi_a$, $\bmtau_a$, $\p(\x_\n, \tv_n)$
		\Statex Definition of averaging and rounding functions:
		\begin{align}
			\myopexpecation_{n|a}\{\p(\X_\n{=}x, \Tstar_n{=}t)\}{=}\frac{1}{|\alpha_\star^{-1}(a)|}\quad\sum_{\mathclap{n\in\alpha_\star^{-1}(a)}}\quad\p(\X_\n{=}x, \Tstar_n{=}t)%
			\label{eq:avgpxt}
		\end{align}
		\begin{equation}
		\oprnd_\kappa(\ell)=\operatorname{sgn}(\ell)\lfloor |\ell| /\kappa+0.5\rfloor\label{eq:symrnd}
		\end{equation}
		\Statex \Comment{design of reconstruction functions}
		\State $\phich_{j}(t)=\oprnd_{\kappav}(L(\B_j|\Tch_j=t))$ using \cref{eq:symrnd}
		\ForAll{$a{\in}\{\alphac(m){:}m{\in} \tilden,n{\in}\setUv\}$}\linelabel{line:startdesingrecvn}%
		\State $\p(\barX=x, \barTc=t)=\myopexpecation_{n|a}\{\p(\X_n=x, \Tc_n=t)\}$ using \cref{eq:avgpxt}
		\State$\phic_{a}(t)=\oprnd_{\kappav}(L(\barTc=t|\barX))$ using \cref{eq:symrnd}\linelabel{line:cnreconstruction}
		\State Compute $\phic_{a}(t)$ $\forall t{\in}\{1,\ldots,2^{w-1}\}$ and store it in $\bmphi_a$.
		\EndFor\linelabel{line:enddesingrecvn}
		\Statex \Comment{design of quantization functions\hfill}
		\ForAll{$a{\in}\{\alphav(n):n\in\setUv\}$}\linelabel{line:loopvnquantdesign}%
		\State $\p(\X_n=x,\Lv_n=\ell)$ with (linearly) scaled version of \cref{eq:vn_computation}: %
		\begin{align}
			\Lv_n=\phich_{\opcol(n)}(\Tch_{\opcol(n)}){+}\sum_{\mathclap{m\in \tilden}}\phic_{\alphac(m)}(\Tc_m)
			\label{eq:nonquantizedvn}
		\end{align}
		\State $\p(\barX=x,\barLv=\ell)=\myopexpecation_{n|a}\{\p(\X_n=x, \Lv_n=\ell)\}$ with \cref{eq:avgpxt} $\forall x,\ell$. %
		\State\linelabel{line:designtau}Design $w$-bit $\quant_a$ maximizing $\I(\barX;\quant_a(\barLv))$. %
		\State\linelabel{line:storetau}Store thresholds $(\tau_{1},\ldots,\tau_{2^{w-1}{-}1})$ of $\quant_a$ in $\bmtau_a$.
		\EndFor\linelabel{line:endloopvnquantdesign}
		\State\linelabel{line:updatedistvtc}Update $p(x_n, t^{\mathsf{v}}_n)$ $\forall n {\in} \setUv$ using \cref{eq:updatedistvtc} with $\star=\labelvn$.
		\begin{align}
			\p(\x_\n, \tstar_n)=\sum_{\lstar_n}\delta(\tstar_\n{-}\quant_{\alphav(n)}(\lstar_n))\p(\x_\n, \lstar_n).\label{eq:updatedistvtc}
		\end{align}\vspace{-0.2cm}
	\end{algorithmic}
\end{algorithm}
\endgroup

\begingroup
\setlength\belowdisplayskip{5pt} %
\setlength\abovedisplayskip{5pt} %
\begin{algorithm}[t]
	\caption{Design of Computational Domain \gls{CN} Update}\label{alg:cndesigncomp}
	\begin{algorithmic}[1]
		\footnotesize
		\Require $w, \setUc,\alphav,\alphac,\p(\x_\n, \tc_\n), \p(\x_\n, \tv_n),\kappac$
		\Ensure $\mytuple{\phi}_a$, $\mytuple{\tau}_a, \p(\x_\n, \tc_n)$
		\Statex \Comment{design of reconstruction functions}
		\ForAll{$a{\in}\{\alphav(m){:}m{\in} \tilden,n{\in}\setUc\}$}
		\State $\p(\barX=x, \barTv=t)=\myopexpecation_{n|a}\{\p(\X_n=x, \Tv_n=t)\}$ using \cref{eq:avgpxt}
		\State $\bar{\varphi}(t)=\neglogtanh(|\opllr(\barTv=t|\barX)|)$ using \cref{eq:neglogtanhfnc} %
		\State$\phiv_{a}(t)=\oprnd_{\kappac}(\bar{\varphi}(t))$ using \cref{eq:symrnd}\linelabel{line:vnreconstruction}
		\State Compute $\phiv_{a}(t)$ $\forall t{\in}\{1,\ldots,2^{w{-}1}\}$ and store it in $\bmphi_a$.
		\EndFor
		\Statex \Comment{design of quantization functions}
		\ForAll{$a{\in}\{\alphac(n):n\in\setUc\}$}\linelabel{line:loopcnquantdesign}
			\State $\p(\X_n=x,\Lc_n=\ell)$ with a (non-linearly) scaled version of \cref{eq:cn_computation}:
			\begin{align}\begin{split}
			\Lc_n=&\prod_{\mathclap{{m\in \tilden}}} \operatorname{sgn}(\Tv_m) \tilde{\varphi}(\sum_{{m\in \tilden}} \phiv_{\alphav(m)}(|\Tv_m|))\\\label{eq:cn_computation_impl2}
			&\text{using }\tilde{\varphi}(\dummyvar)=\dummyvar_{\mathsf{max}}{-}\dummyvar,\quad \dummyvar{\in}\{0,1,\ldots,\dummyvar_{\mathsf{max}}{-}1\}
		\end{split}\end{align}
			\State $\p(\barX=x,\barLc=\ell)=\myopexpecation_{n|a}\{\p(\X_n=x, \Lc_n=\ell)\}$ with \cref{eq:avgpxt} $\forall x,\ell$. %
		\State Design $w$-bit $\quant_a$ maximizing $\I(\barX;\quant_a(\barLc))$.
	\State Store thresholds $(\tau_{1},\ldots,\tau_{2^{w-1}{-}1})$ of $\quant_a$ in $\bmtau_a$.
	\EndFor\linelabel{line:endloopcnquantdesign}
	\State Update $\p(\x_\n, \tc_n)$ $\forall n {\in} \setUc$ using \cref{eq:updatedistvtc} with $\star=\labelcn$.
\end{algorithmic}
\end{algorithm}
\endgroup

\begingroup
\setlength\belowdisplayskip{0pt}
\begin{algorithm}[t]
	\caption{Design of min-sum \gls{CN} update.}\label{alg:cndesignminsum}
	\begin{algorithmic}[1]
		\footnotesize
		\Require $\setUc,\p(\x_\n, \tv_n)$
		\Ensure $\p(\x_\n, \tc_\n)$
		\State $\forall n {\in} \setUc{:}$ Update distribution of \gls{CN}-to-\gls{VN} messages:
		\begin{align}
		\p(\x_\n, \tc_\n)=\sum_{\mathclap{\substack{\myvec{x}=(x_m:m{\in} \tilden):\\
		\x_\n=\bigoplus(\myvec{x})}}}\quad\,\quad\sum_{{\substack{\myvec{t}^\labelvn=(\tv_m:m{\in} \tilden):\\
					\tc_n=\operatorname{MS}(\myvec{t}^\labelvn)}}}\prod_{m\in \tilden} \p(x_{m}, \tv_{m})
					\label{eq:ddeminsum}
					\end{align}
		where $\operatorname{MS}$ is defined in \cref{eq:cn_computation_min_approx}.
				\end{algorithmic}
\end{algorithm}
\endgroup

The decoding performance of the alignment strategies is compared in this section using a 5G LDPC code, \gls{BPSK} modulation and an \gls{AWGN} channel.

The channel messages are quantized with $\wch=4$\,bits for all $w$-bit decoders.
The distribution $\p(\breal_j,\tch_j)$ depends on the design-$\EbNo$ which we optimize to minimize \gls{FER} of the designed decoder (see \cref{alg:chdes,alg:chdist}).
A flooding schedule performs a maximum of $\itermax=30$ decoder iterations.
The decoding is stopped earlier when all core parity checks are satisfied.
Core parity checks correspond to \glspl{CN} that involve only inputs from \glspl{VN} with $\dv_j{>}1$~\cite{frenzel19}.
A frame error is counted if any of the information bits or core parity bits is falsely decoded.

\begin{figure}[t]
	\centering
	\includestandalone[mode=\mymodestandalone]{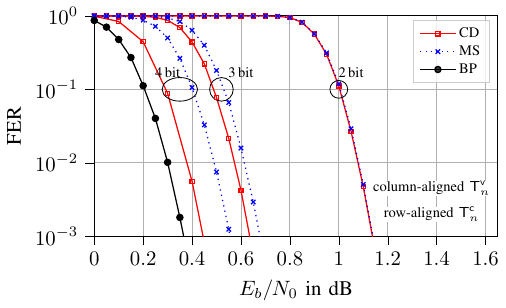}
	\vspace{-0.2cm}
	\caption{Performance comparison of the computational domain (CD) \gls{CN} update (cf. Alg.\,\ref{alg:cndesigncomp}) and the min-sum (MS) \gls{CN} update (cf. Alg.\,\ref{alg:cndesignminsum}). Under 2-bit decoding both node updates deliver equivalent performance (flooding schedule, $\itermax=30$, $r=\nicefrac{1}{3}$, $K=8448$). }
	\label{fig:fer_alignment_ms_vs_cd}
\end{figure}

In \cref{fig:fer_alignment_ms_vs_cd} we compare the \gls{FER} performance of the computational domain \gls{CN}~\cref{eq:cn_computation} and the min-sum \gls{CN}~\cref{eq:cn_computation_min_approx}.
It can be observed that the computational domain update shows increasing gains for higher message resolutions.
For 2-bit decoding, the performance is almost the same, since the error from coarse quantization is much larger than the error from using the min-sum approximation \cref{eq:cn_computation_min_approx}.
Compared to a high-resolution \gls{BP} algorithm~\cite{gal62}, the computational domain decoders show a performance loss of 0.07\,dB, 0.27\,dB and 0.77\,dB for 4-bit, 3-bit, and 2-bit decoding, respectively.

\begin{figure}[t]
	\centering
	\includestandalone[mode=\mymodestandalone]{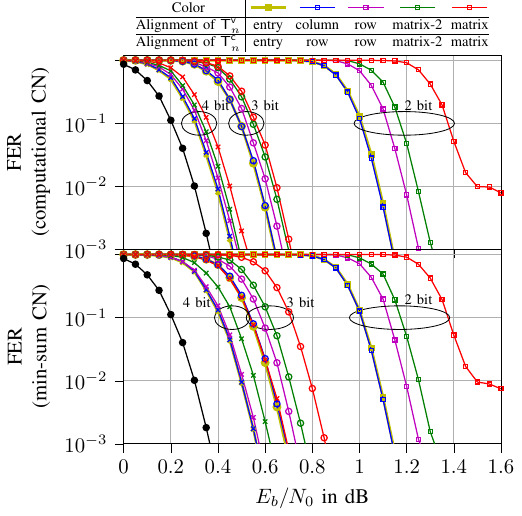}
	\vspace{-0.8cm}
	\caption{The different alignment strategies from \cref{fig:alignment_regions} lead to significant differences in performance (flooding schedule, $\itermax=30$, $r=\nicefrac{1}{3}$, $\myK=8448$). The results show that the configuration of the alignment regions is more important at lower resolutions. The matrix alignment under the flooding schedule corresponds to the design proposed in \cite{kang22}.}
	\vspace{-0.0cm}
	\label{fig:fer_alignment}
\end{figure}

The performance of the individual alignment strategies for the two update variants is shown in \cref{fig:fer_alignment}.
Overall, the performance degradation from using inferior alignments is more noticeable at lower resolutions.
For example, under 2-bit decoding the column-row alignment outperforms the matrix-matrix alignment by 0.4~dB.

The entry alignment (cf. \cref{fig:alignment_regions}a) achieves the best performance but causes the highest complexity in design and implementation since every quantizer and reconstruction operation is designed individually for each memory location $\n$.

The column-row alignment (cf. \cref{fig:alignment_regions}c\,\&\, \cref{fig:alignment_regions}b) saves complexity by designing common quantizers for all memory locations in each column or row, respectively.
Thus, the quantization is specifically designed for each node degree.

It can be observed in \cref{fig:fer_alignment}, that designing degree-specific quantization and reconstruction functions achieves similar performance as the entry-entry configuration.

The row-row alignment (cf. \cref{fig:alignment_regions}b) forces all inputs and outputs of a \gls{CN} to be represented by the same random variable.
As all quantized inputs represent the same LLR alphabet, the LLR reconstruction in the min-sum update \cref{eq:cn_computation_min} can be avoided by design.
Still, in \cref{fig:fer_alignment}, superior performance of the column-row alignment over the row-row alignment can be observed.
In both cases the min-sum rule \cref{eq:cn_computation_min_approx} is used where no LLR translation takes place.
One explanation could be that the \gls{VN} quantizer threshold design of this section is not aware of the subsequent \gls{CN} processing, leading to sub-optimal performance as shown in the next \cref{sec:cnaware_quantization}.

The matrix alignment (cf. \cref{fig:alignment_regions}e) yields a design with the same reconstruction functions and quantization thresholds for all memory locations in one update step.
However, a significant performance degradation is observed.
Particularly, the 2-bit decoding shows a relatively high error floor.
The error floor can be significantly reduced using the matrix-2 alignment (cf. \cref{fig:alignment_regions}d).
This strategy comprises two complementary regions where the first region excludes rows connected to degree-one \glspl{VN}.
In this way, \gls{CN} messages within the first region are not limited by the degree-one \gls{VN} reliability improving the performance particularly under 2-bit decoding.

\subsection{Comparison to Two-Input Lookup Table Decoders}\label{subsec:comparison-to-two-input-lookup-table-decoders}
\begin{figure}[t]
	\includestandalone[mode=\mymodestandalone]{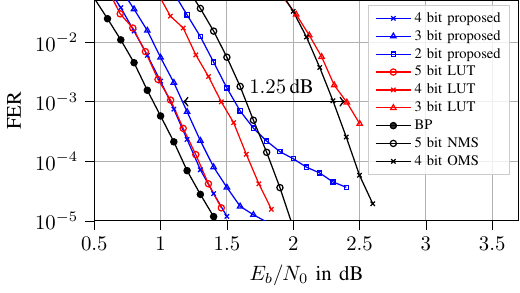}
	\vspace{-0.8cm}
	\caption{Performance comparison to \gls{LUT}-based decoders \cite{stark20} and min-sum decoders\cite{chen05} (5-bit \gls{NMS} and 4-bit \gls{OMS}) under a flooding schedule, $\itermax=100$, $r=\nicefrac{1}{3}$, $\myK=1032$ and 5G-like LDPC code from \cite{stark20}.
		The proposed decoder uses column-row alignment (cf. \cref{fig:alignment_regions}c\,\&\,\cref{fig:alignment_regions}b), a computational domain \gls{VN} update and min-sum \gls{CN} update.}
\vspace{-0.0cm}
\label{fig:fer_alignment_vs_conventional}
\end{figure}
In~\cite{stark20} LUT-based decoders have been proposed for 5G-like LDPC codes.
All node operations are realized with concatenations of two-input LUTs.
For example, a \mbox{degree-$d$} \gls{VN} with underlying bit $\x$ observes the quantized extrinsic messages $\myvec{y}=(y_1,\ldots,y_{d})$ that shall be mapped to a compressed message $t$.
As described in \cref{sec:vndesign}, a mutual information maximizing node operation performs $t=Q(\opllr(x|\myvec{y}))$.
Implementing this mapping with a single lookup $t=\operatorname{LUT}(\myvec{y})$ causes high complexity due to many possible input combinations.
Hence,~\cite{stark20} considers a concatenation of two-input LUTs that sequentially generate the output message $t=t_{d}$ with $t_{k}=\operatorname{LUT}_k(t_{k-1},y_{k})=Q_k(\opllr(x|t_{k-1},y_{k}))$ where $t_{1}=y_1$ and $k{\in}\{2,{..},d\}$.
The multiple compression steps $Q_k$ cause additional mutual information loss compared to a single quantization step, also mentioned in~\cite{he19}.
This work avoids concatenated compression steps by computing $\opllr(x|\myvec{y})$ with high-resolution arithmetic operations followed by threshold quantization $Q$.

It can be observed in \cref{fig:fer_alignment_vs_conventional} that the proposed decoders use fewer bits for the exchanged messages to achieve similar or even better performance.
For example, the proposed \mbox{4-bit} decoder achieves similar performance as the 5-bit \gls{LUT} decoder and outperforms the 4-bit \gls{LUT} decoder by 0.39\,dB\@.
The proposed 3-bit decoder even outperforms the 4-bit \gls{LUT} decoder by 0.29\,dB and the 3-bit \gls{LUT} decoder by 1.25\,dB\@.
The proposed 2-bit decoder shows an error floor below $\mathrm{FER}=3{\times}10^{-4}$.
However, above the error floor, it still manages to operate within 0.1\,dB w.r.t.\ the 4-bit \gls{LUT} decoder.

From the results, it can be concluded that the \gls{LUT}-based decoders suffer severely from multiple intermediate quantization steps when using low code rates and when the internal bit width is below 5 bits.
We remark that the decoders in~\cite{stark20} used a quantizer design that realizes a matrix alignment (cf. \cref{fig:alignment_regions}e).
Furthermore, all proposed decoders use 4 bits for the channel message.
The proposed decoders can achieve excellent performance even at 2-bit quantization under a column-row alignment.

\subsection{Comparison to Conventional Min-Sum Decoders}\label{subsec:comparison-to-conventional-min-sum-decoders}
In~\cite{chen05} reduced complexity \gls{LLR}-based decoding algorithms were developed which make use of approximations to simplify the design and implementation of the min-sum decoder.
Two commonly used variants are known as the \glsreset{NMS}\gls{NMS} or \glsreset{OMS}\gls{OMS} algorithm.
One of the main differences compared to the proposed algorithms is that the quantization levels are not adjusted as the number of iterations increases.
Hence, a higher bit width is required to offer a sufficient representation range for the exchanged LLRs. Using only 4 bits leads to significant performance degradation which can be observed in \cref{fig:fer_alignment_vs_conventional}.
The min-sum decoders were configured as in~\cite{stark20}.
The 4-bit quantized \gls{OMS} decoder is outperformed by 1.2\,dB compared to the proposed 4-bit decoder.
Using a 5-bit quantized \gls{NMS} decoder reduces the difference to 0.7\,dB\@.

\section{Check-Node-Aware Design of Non-Uniform Quantization for the Variable Node Update}\label{sec:cnaware_quantization}
\begin{figure}[t]
	\centering
	\vspace{-0.1cm}
	\includegraphics[width=0.95\linewidth]{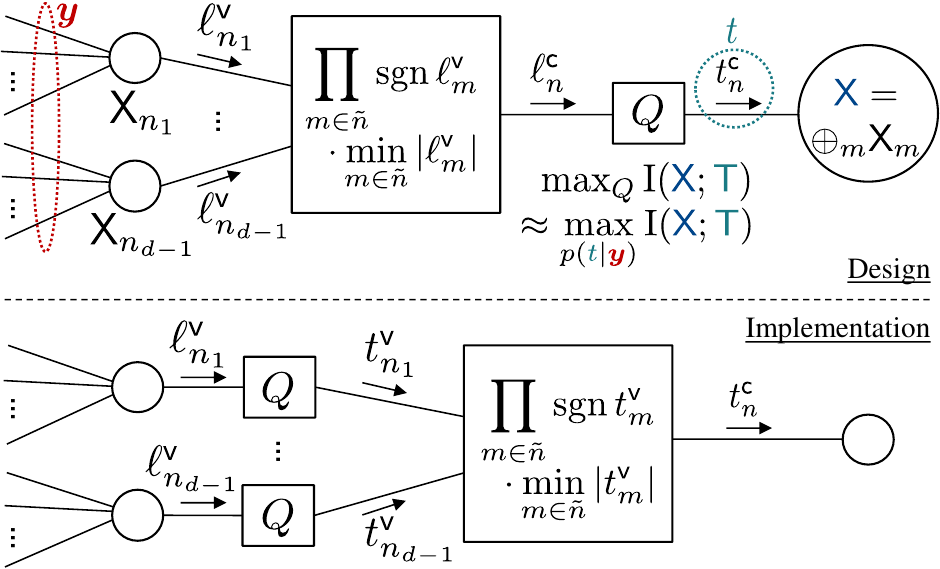}
	\vspace{-0.0cm}
	\caption{A \gls{CN}-aware \gls{VN} quantizer design extends the \gls{IB} setup from \cref{fig:chvncn} optimizing the reliability levels of \gls{CN} messages. Without performance loss, the quantizer designed after the \gls{CN} update can be moved before the \gls{CN} to reduce complexity.}
	\vspace{-0.0cm}
	\label{fig:cnaware}
\end{figure}
This section shows that the mutual information between \gls{CN} messages and code bits, $\I(\X_\n;\Tc_\n)$, is severely influenced by the choice of the quantization levels of the \gls{VN} quantizer $Q$.
Based on the results of our work~\cite{mohr22aware} this section proposes an extended \gls{IB} setup to design VN quantizers aiming at maximizing mutual information~$\I(\X_\n;\Tc_\n)$.

Fig.~\ref{fig:cnaware} depicts a \gls{CN} update using \emph{non-quantized} messages $\lv_{n_k}$ from connected extrinsic \glspl{VN} to obtain a \gls{CN} message~$\tc_\n$ for a target \gls{VN} with bit $\x_\n=\oplus_{m\in \tilden}\x_{m}$.
The inputs to the extrinsic \glspl{VN} are denoted as $\myvec{\iby}$.
The messages $\x_\n,\myvec{y}$ and $\tc_\n$ form a \gls{CN}-aware IB setup with the corresponding relevant, observed, and compressed variables $\X$, $\myvec{\myrv{Y}}$, and $\myrv{T}$, respectively.
The following derives a computationally efficient solution for realizing a mapping $p(t|\myvec{y})$ aiming for $\max_{p(t|\myvec{y})}\I(\X;\myrv{T})$.
For the design we postpone the quantization at the \gls{VN}.
Thus, a \gls{CN} update is carried out for all \gls{CN} memory locations $n{\in}\mathcal{N}$ using the min-sum operation \cref{eq:cn_computation_min} with non-quantized \gls{VN} outputs~$\lv_\n$ according to \cref{eq:nonquantizedvn} (cf. \cref{line:pxlv_cnaware,line:pxlc_cnaware} of \cref{alg:vndesignaware}).
Then, the quantizer operation leads to the output messages
\begin{align}
	\tc_\n = \quant(\lc_\n) &= \prod_{m\in n}\opsgn(\lv_{m}) \quant(\min_{m\in n} |\lv_{m}|).
	\label{eq:lookahead_quant_after}
\end{align}
The quantizer function $Q\,{:}\, \mathcal{L}^\labelcn_\n{\rightarrow}\mathcal{T}^\labelcn_\n$ is defined by a set of symmetric quantizer thresholds~$\myvec{\tau}$ as in \cref{sec:mim_channel_quant}.

For applying the alignment regions from section \ref{sec:message_alignment_regions}, we model the non-quantized output $\lc_\n$ using the random variable $\Lc_\n$.
With mixture variables $\bar{\X}_a$ and $\barLc_a$ an aligned quantizer design can  be performed with objective $\max_{Q_a}\I(\barX_a;\barTc_a)$ (cf. \cref{line:pavgxl,line:mimax_cnaware} of \cref{alg:vndesignaware}), resulting in a specific quantizer $Q_a$ for each alignment region $a{\in}\setAc$ of the \gls{CN} memory.
The row, matrix-2, or matrix alignment are suitable choices as all inputs to one \gls{CN} use the same quantizer.

As shown in \cref{fig:cnaware}, the resulting quantizer can equivalently be applied before the min-sum \gls{CN} update. This approach significantly reduces the number of exchanged bits and simplifies the internal \gls{CN} update complexity.

We remark that the min-sum output $\ell^\labelcn_\n$ only approximates the \gls{LLR} value $\opllr(\x_\n|\myvec{y})$, where $\myvec{y}$ denotes the inputs to connected extrinsic \glspl{VN} (cf. \cref{fig:cnaware}). Thus, threshold quantization applied to $\ell^\labelcn_\n$ approximates the information-optimal compression operation $\p(\ibt|\myvec{\iby})$. Nevertheless, this design method achieves excellent performance improvements.

\subsection{Analysis of Quantizer Thresholds}
\begin{figure}[t]
	\includestandalone[mode=\mymodestandalone]{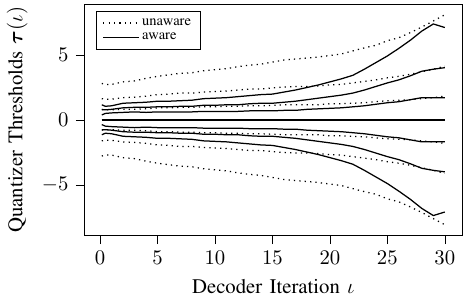}
	\vspace{-0.2cm}
	\caption{Threshold placement of $Q_{a=1}$ for every decoder iteration (row alignment for \gls{VN} and \gls{CN} memory, flooding schedule, $K=8448$, $r=\nicefrac{1}{3}$).}
	\vspace{-0.0cm}
	\label{fig:cn_aware_boundary_evolution}
\end{figure}

\begin{figure}[t]
	\includestandalone[mode=\mymodestandalone]{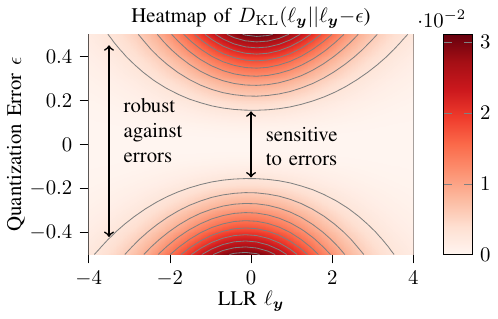}
	\vspace{-0.3cm}
	\caption{The heatmap shows that a quantization error $\epsilon=\ell_{{\myvec{\iby}}}{-}\ell_{t=Q(\ell_{\myvec{\iby}})}$ causes higher loss of mutual information if $\ell_{\myvec{\iby}}$ is unreliable, i.e., near zero.
		}
	\label{fig:cn_aware_kl_divergence}
\end{figure}

\cref{fig:cn_aware_boundary_evolution} depicts quantizer thresholds $\myvec{\tau}(\iota)$ for every decoder iteration~$\iota$. It can be observed that the \gls{CN}-aware design method results in a denser placement of thresholds around the decision boundary for most iterations. This behavior can be explained as follows: consider $\ell_{\myvec{\iby}}=L(x|\myvec{\iby})$ and $t$ as the input and output of a quantizer $\quant$, respectively.
The optimization of the quantizer aims to preserve information about a bit $\ibxreal$ by minimizing a mutual information loss
\begin{align}
	\I(\ibX;\myvec{\ibY}){-}\I(\ibX;\ibT)=\sum_{\myvec{\iby}} p(\myvec{\iby}) D_{\mathrm{KL}}(\ell_{\myvec{y}}||\ell_{t=Q(\ell_{\myvec{y}})})
	\label{eq:quan_mi_loss}
\end{align}
with $D_{\mathrm{KL}}(\ell_{\myvec{\iby}}||\ell_t)=\sum_{\ibx} p(x|\myvec{\iby})\log \frac{p(\ibx|\ell_{\myvec{\iby}})}{p(\ibx|\ell_\ibt)}$ and $p(\ibx|\ell)=\allowbreak\frac{e^{\ell(1-\ibx)}}{1+e^{\ell}}$. 
In \cref{fig:cn_aware_kl_divergence} the Kullback-Leibler divergence $D_{\mathrm{KL}}$ (always non-negative) is shown in relation to the quantization error $\epsilon=\ell_{\myvec{\iby}}{-}\ell_\ibt$ and input LLR level $\ell_{\myvec{\iby}}$.
The contour lines reveal that the mutual information loss is more sensitive to quantization errors when the inputs are unreliable.
The fraction of unreliable messages is much larger at the \gls{CN} output $\lc_\n$ than at the \gls{CN} input $\lv_\n$ considering the min-sum update~\cref{eq:lookahead_quant_after}. Hence, the \gls{CN}-aware thresholds are placed more densely close to the decision threshold to reduce quantization errors $\epsilon$.

\subsection{Comparison to \gls{CN}-Unaware Optimization}
\begin{figure}[t]
	\centering
	\includestandalone[mode=\mymodestandalone]{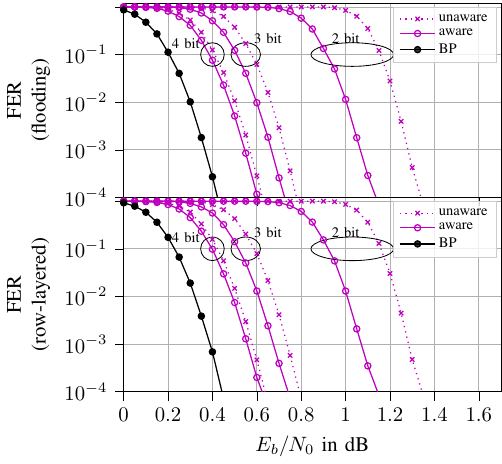}
	\vspace{-0.4cm}
	\caption{Performance with and without \gls{CN} awareness (flooding schedule $\itermax=30$ or layered schedule $\itermax=15$, $r=\nicefrac{1}{3}$, $\myK=8448$, computational domain \gls{VN}, min-sum \gls{CN}, row alignment).}
	\vspace{-0.3cm}
\label{fig:fer_cn_aware_ms}
\end{figure}

\begin{algorithm}[t]
	\caption{Design of \gls{CN}-Aware \gls{VN} Update}\label{alg:vndesignaware}
	\begin{algorithmic}[1]
		\footnotesize
		\Require $w$, $\setUv$, $\alphac$, $p(\breal_j,\tch_j)$,  $p(\x_\n, \tc_\n)$, $p(\x_\n, \tv_\n)$
		\Ensure $\bmphi_a$, $\bmtau_a$, $p(\x_\n, \tv_\n)$
		\State Design reconstructions $\bmphi_a$ as in~\crefrange{line:startdesingrecvn}{line:enddesingrecvn} of~\cref{alg:vndesign}.
		\State Create set $\mathcal{U}_{\mylabel{rows}}=\{n{:} \operatorname{row}(n){\in}\{\operatorname{row}(n'),n'{\in}\setUv\}\}$.%
		\State \linelabel{line:pxlv_cnaware}Compute $p(\x_\n, \lv_\n)$ $\forall n {\in }\mathcal{U}_{\mylabel{rows}}$ with $\lv_\n$ defined in \cref{eq:nonquantizedvn}. %
		\State \linelabel{line:pxlc_cnaware}Compute $p(\x_\n, \lc_\n)$ $\forall n {\in}\mathcal{U}_{\mylabel{rows}}$ with\,\cref{eq:ddeminsum} using $\lv_\n$ and $\lc_\n$ instead of $\tv_\n$ and $\tc_\n$, respectively.
		\For{$a{\in}\{\alphac(n){:}n{\in}\mathcal{U}_{\mylabel{rows}}\}$}
		\State\linelabel{line:pavgxl} $p(\barX=x,\barLc=\ell)=E_{n|a}\{p(\X_\n=x, \Lc_\n=\ell)\}$ with \cref{eq:avgpxt} $\forall x,\ell$
		\State\linelabel{line:mimax_cnaware}\linelabel{line:qcnaware}Design $\quant_a$ maximizing $\I(\barX;\quant_a(\bar{\myrv{L}}^\labelcn))$.
		\State Store thresholds $(\tau_{1},\ldots,\tau_{2^{w-1}{-}1})$ of $\quant_a$ in $\bmtau_a$. %
		\EndFor
		\State Update $p(\x_\n, \tv_\n)$ $\forall n{\in}\setUv$ using \cref{eq:updatedistvtc} with $\star=\mylabel{v}$.
	\end{algorithmic}
\end{algorithm}
\cref{fig:fer_cn_aware_ms} compares the FER performance when using the \gls{CN}-aware and -unaware design method.
In the upper part of \cref{fig:fer_cn_aware_ms} a flooding schedule is used.
\gls{CN} awareness improves the performance by 0.03, 0.06, and 0.2\,dB when using 4,\,3, or 2\,bits for the exchanged messages, respectively.
In the lower part of \cref{fig:fer_cn_aware_ms} a row-layered schedule is used.
Although the iteration count is limited to $\itermax=15$ similar performance as with the flooding schedule where $\itermax=30$ is achieved.
Also, the performance gains of the \gls{CN}-aware design technique are similar.
From the results, it can be concluded that extending the optimization scope to include \gls{CN} outputs can improve performance, particularly at low resolutions (e.g., 2 or 3 bits), without increasing node update complexity.

\subsection{Comparison to Neural Network Finite Alphabet Decoders} %
\begin{figure}[t]
	\includestandalone[mode=\mymodestandalone]{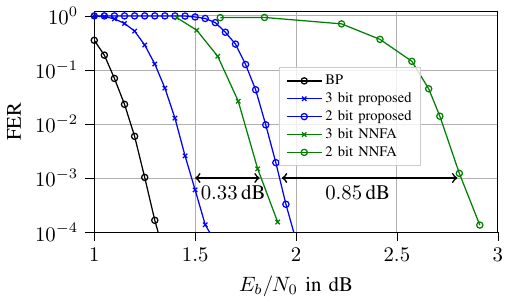}
	\vspace{-0.4cm}
	\caption{Comparison to neural network trained finite alphabet (NNFA) decoders\cite{lyu23} ($r=\nicefrac{1}{2}$, $K=8448, \itermax=20$). The proposed decoders use \glspl{VN} and min-sum \glspl{CN} with column-row alignment.}
	\vspace{-0.1cm}
	\label{fig:literature_comparison_to_lyu23}
\end{figure}
In~\cite{kumar13,ly17} so-called non-surjective finite-alphabet decoders were proposed.
These decoders compute finite alphabet \gls{CN} and \gls{VN} messages as
$\tc_\n=\prod_{m\in \tilden} \operatorname{sgn}(\tv_{m})\min_{m\in \tilden}(|\tv_\n|)$ and  $\tv_\n=\quant(\phich(\tch_j){+}\sum_{m\in \tilden}\phi(\tc_m))$, respectively.
The quantization $\quant{:}\mathcal{L}{\to}\mathcal{T}^\labelvn$ is classified as a non-surjective function and the reconstruction of \gls{CN} messages is $\phi{:} \mathcal{T}^\labelcn{\to}\mathcal{L}$.
While the structure is similar to our work, the quantization $\quant$ and reconstruction $\phi$ are designed differently.
For example \cite{ly17} used heuristic techniques to optimize $\quant$ and $\phi$ for irregular codes.
In \cite{lyu23} the design was improved by training $\quant$ and $\phi$ with a recurrent quantized neural network model for every two iterations of a flooding schedule.
In \cref{fig:literature_comparison_to_lyu23} our decoders outperform the neural network finite alphabet (NNFA) decoders~\cite{lyu23} by up to $0.85$\,dB at 2-bit resolution.

One reason for the better performance might be the usage of a probability-based design instead of a data-driven design.
We optimize each $\quant$ with an IB algorithm based on probability distributions tracked with discrete density evolution.
Also the reconstruction levels $\phi$ are accurately obtained from the tracked probabilities instead of training them.
Furthermore, we design individual $\phi$ and $\quant$ per iteration for every row of the base matrix with a row alignment.
The NNFA decoders use the same $\phi$ and $\quant$ for every two iterations.

\section{Schedule Optimization for 5G LDPC Codes under Quantized Decoding}\label{sec:layeredopt}
\begin{figure}[t]
	\centering
	\includestandalone[mode=\mymodestandalone]{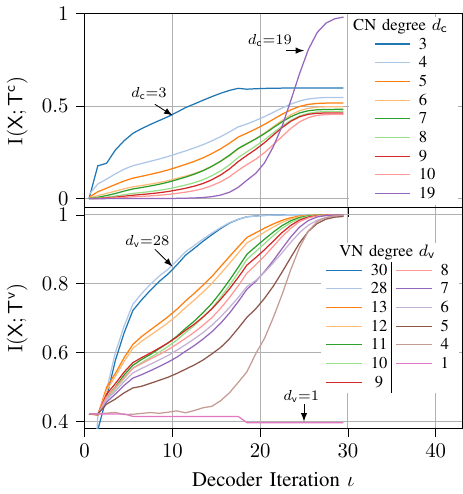}
	\vspace{-0.2cm}
	\caption{Degree-specific mutual information $\I(\X;\Tc)$ and $\I(\X;\Tc)$ for \gls{CN} messages and \gls{VN} messages, where $\X$ and $\Tstar$,  $\star{\in}\{\labelcn,\labelvn\}$, are defined by
		$p(\X=x;\Tstar=t)=E_{n|d_\star}\{p(\X_n=x,\Tstar=t)\}$
		}%
	\vspace{-0.0cm}
	\label{fig:mi_evolution_degree}
\end{figure}
In layered decoding, the layers typically form groups of nodes with specific degrees.
\cref{fig:mi_evolution_degree} evaluates the degree-specific mutual information between code bits and exchanged messages under a flooding schedule, showing different slopes across the iterations.
This observation motivates optimizing the schedule to prioritize layers with the highest incremental mutual information gains. Such optimization is particularly beneficial when node degrees exhibit significant disparity. Unlike dynamic schedules~\cite{wang_two_2020,ren24}, the proposed approach produces a static schedule, avoiding real-time calculations.

\begin{figure}[t]
	\centering
	\resizebox{0.99\linewidth}{!}{\includestandalone[mode=\mymodestandalone]{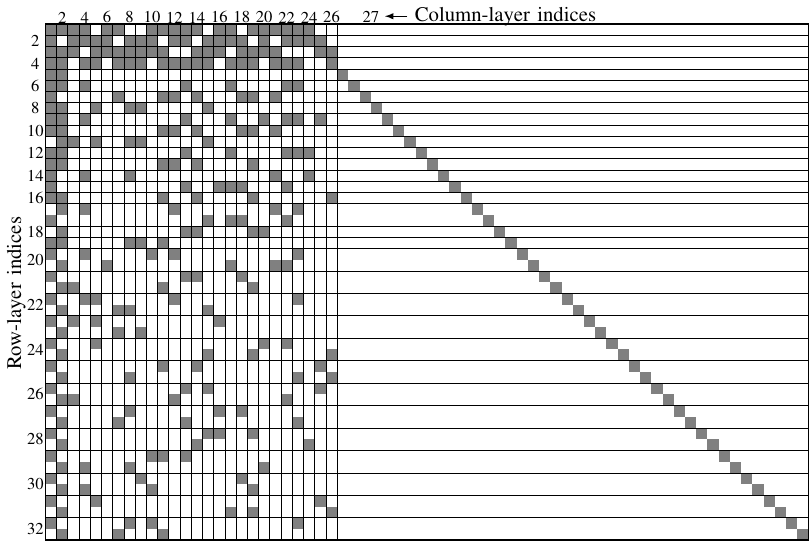}}
	\vspace{-0.5cm}
	\caption{Illustration of layers consisting of orthogonal rows and columns.}
	\vspace{-0.0cm}
	\label{fig:orthogonal_rows_cols}
\end{figure}
\vspace{-0.1cm}
\subsection{Row-Layered Schedule Optimization}
\begin{algorithm}[t]
	\caption{Design of Decoder With Optimized Row-Layers}\label{alg:decdesignopt}
	\begin{algorithmic}[1]
		\footnotesize
		\Require $w, \Uinit,\alphac,\alphav, p(\breal_j,\tch_j)$
		\Ensure $\mytuple{\myset{U}}$, $\mytuple{\phi}_k$, $\mytuple{\tau}_k$		
		\State Run \cref{alg:decdesign} with initialization schedule  $\Uinit$.
		\State $k=|\Uinit|$
		\While{$\sum_{k'=0}^{k-1} |\mathcal{U}_{k'}^\star|/2|\mathcal{N}| < \itermax$}
		\State $\forall \n{\in} \setN$: Compute $\p(\x_\n,\lv_\n)$ with $\lv_\n$ defined in \cref{eq:nonquantizedvn}.
		\State $\forall n{\in} \setN$: Compute $\p(\x_\n,\lc_\n)$ using \cref{eq:ddeminsum} with $\lv_\n$ and $\lc_\n$ instead of $\tv_\n$ and $\tc_\n$, respectively.
		\State Store layer $l$ optimizing $\DeltaIc_l$ in \cref{eq:mi_gain_row_layered} as $\setUv_k=\mathcal{M}_l$ and $\setUc_{k+1}=\mathcal{M}_l$.
		\State $\mytuple{\phi}_k,\mytuple{\tau}_k \gets$ Design update $\setUv_k$ with \cref{alg:vndesign} or \cref{alg:vndesignaware}.
		\State $\mytuple{\phi}_{k+1},\mytuple{\tau}_{k+1} \gets$ Design update  $\setUc_{k+1}$ with \cref{alg:cndesigncomp} or \cref{alg:cndesignminsum}.
		\State $k=k+2$
		\EndWhile
	\end{algorithmic}
\end{algorithm}
A row-layered schedule divides a base matrix into $32$ layers as enumerated in \cref{fig:orthogonal_rows_cols} for the base graph 1. Some layers comprise multiple rows. These rows represent CNs without common VNs in their neighborhood and can be updated simultaneously without degrading decoding performance~\cite{cui_design_2021}.
A single layer update comprises \gls{VN} updates~$\setUv$ followed by \gls{CN} updates~$\setUc$ pointing to the memory locations $\mathcal{M}_l$ in layer $l$.
Typically, the updates are performed one after another (cf.~ \cref{fig:orthogonal_rows_cols}).
To accelerate the convergence, it is suggested to select the layer $l_\mathsf{opt}=\arg \max_l \DeltaIc(l)$ in each update step which provides the highest mutual information gain
\begin{align}
	\DeltaIc(l)=\sum_{\mathclap{n\in \mathcal{M}_l}}(\I(\X_\n;{\uLc_n}){-}\I(\X_\n;\Lc_n)/|\mathcal{M}_l|.\label{eq:mi_gain_row_layered}
\end{align}
In \cref{eq:mi_gain_row_layered} the variables $\Lc_n$ are the non-quantized \gls{CN} outputs as defined in~\eqref{eq:lookahead_quant_after}.
$\uLc_n$ denotes that the variable $\myrv{L}^c_n$ has been improved through the update of layer $l$.
Using a quantized message would require temporary quantizer designs for every layer which increases design complexity and did not show performance improvements.
The schedule construction finishes when reaching the maximum number of iterations $\itermax$.
\cref{alg:decdesignopt} describes the modified decoder design. 
\subsection{Column-Layered Schedule Optimization}
A column-layered schedule divides the base matrix into $27$ vertical layers as enumerated in \cref{fig:orthogonal_rows_cols} for base graph~1 by merging degree-1 columns to one layer.
A single layer update comprises \gls{CN} updates $\setUc$ followed by \gls{VN} updates $\setUv$ pointing to the memory locations $\mathcal{M}_l$ in layer $l$.
It is proposed to select the layer $l_\mathsf{opt}=\arg \max_l \DeltaIv(l)$ in each layer update step which provides the highest mutual information gain
\begin{align}
	\DeltaIv(l)=\sum_{\mathclap{n\in\mathcal{M}_l}}(\I(\X_n;{\uLv_n}){-}\I(\X_n;\Lv_n))/|\mathcal{M}_l|
\end{align}
where $\uLv_n$ denotes that the \gls{VN}-to-\gls{CN} message variable $\Lv_n$ has been improved with the update of layer $l$.
The realizations of non-quantized \gls{VN} outputs $\Lv_n$ are defined in~\cref{eq:nonquantizedvn}.

\begin{figure}[t]
	\centering
	\includestandalone[mode=\mymodestandalone]{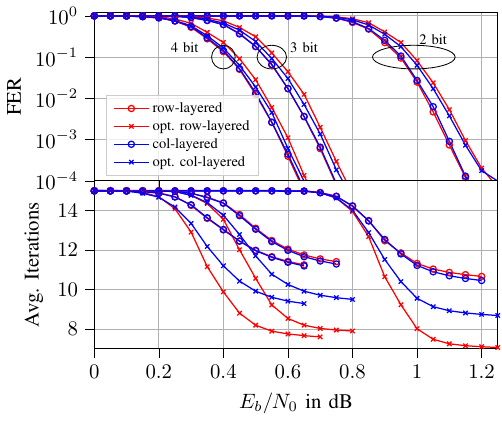}
	\vspace{-0.2cm}
	\caption{Performance of optimized layered decoding schedules ($r=\nicefrac{1}{3}$, $\myK=8448, \itermax=15$, computational domain \gls{VN}, min-sum \gls{CN}, \gls{CN}-aware design with row alignment, same design-$\EbNo$ for each bit width $w$).}
	\vspace{-0.1cm}
	\label{fig:fer_opt_schedules}
\end{figure}

\subsection{Performance of Optimized Schedules}\label{sec:perfoptsched}
\cref{fig:fer_opt_schedules} evaluates the performance when using the standard and optimized layered decoding schedules.
It can be observed that the standard row- and column-layered schedules deliver similar performance in terms of FER and average iteration count.
The average iteration count is proportional to the average number of exchanged messages for successful decoding and significantly influencing area efficiency of decoders with early termination detection.
The optimized column-layered schedule reduces the average iteration count for all bit widths by approximately 20\%. 
The optimized row-layered schedule achieves a reduction of 35\%. 
A 0.03-0.05\,dB performance loss is observed compared to the standard schedule.
The optimization can lead to a non-uniform number of updates of the individual layers, possibly creating more harmful cycle effects that are neglected in the probabilistic decoder model of the design phase.

\section{Design with Rate Compatibility}\label{sec:ratecompatibility}
\begin{figure}[t]
		\centering
				\subfloat[\centering Base graph 1]{\resizebox{0.52\linewidth}{!}{\includestandalone[mode=\mymodestandalone]{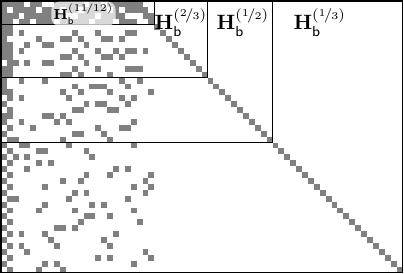}}}
		\hfil
		\subfloat[\centering Base graph 2]{\resizebox{0.45\linewidth}{!}{\includestandalone[mode=\mymodestandalone]{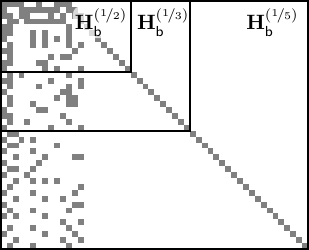}}}
	\caption{Matrices for different rates from base graph 1 or base graph 2.}
	\vspace{-0.0cm}
	\label{fig:ratecompatiblebg1bg2}
\end{figure}

Adapting the code rate is essential in 5G NR for reliable and efficient transmission under varying channel conditions. 5G LDPC codes with different rates are derived from a base matrix, as shown in \cref{fig:ratecompatiblebg1bg2}. Instead of using multiple rate-specific decoders, employing a single rate-compatible decoder reduces receiver complexity. This section presents a rate-compatible decoder for the new proposed techniques, which include alignment regions, \gls{CN}-aware quantization, and layered schedules. The rate-compatible design ensures consistent reconstruction and quantizer parameters for all code rates in each update of a decoding schedule $\mytuple{\myset{U}}$.
\begin{figure*}[t]
	\centering
	\includestandalone[mode=\mymodestandalone]{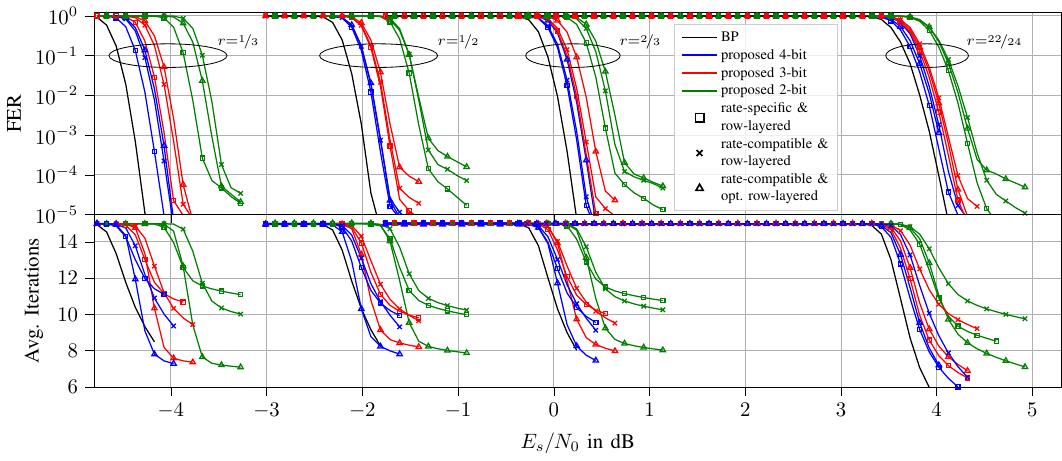}
		\vspace{-0.4cm}
		\caption{Results for rate-compatible decoding ($\myK=8448$, $\itermax=15$, computational domain VN, min-sum CN, \gls{CN}-aware design with row-alignment).}%
		\label{fig:rate_compatible_2bit_layer_opt_er}
	\end{figure*}
\subsection{Modification of the Design Process}
Multiple design code rates are selected with $\myr\in\mathcal{R}$ either for base graph 1 or base graph 2.
Each $r$ identifies a parity-check matrix $\matHbr$, depicted in \cref{fig:ratecompatiblebg1bg2}.
As introduced in \cref{sec:enc_dec_5g}, the non-zero entries of each $\matHbr$ can be seen as memory locations $n^{(r)}{\in}\setNr$ for the exchanged messages in a decoder.
Every code can be decoded with the lowest-rate decoder by deactivating parts of the lowest-rate matrix $\matHb^{(r_0)}$. %

Multiple decoders are designed jointly with the previously described techniques by extending the memory location index~$n$ with a rate dimension such that $n{\in} \mathcal{N}=\{(n^{(r)},r): n^{(r)}{\in}\mathcal{N}^{(r)},r{\in}\mathcal{R}\}$.
Thus, each $n$ identifies a memory location as well as the code rate.
A rate-compatible design follows straightforwardly from the alignment approach in \cref{sec:message_alignment_regions}.

For each code rate $r$ an individual design-$E_b/N_0$ must be specified, leading to individual channel distributions $p(\breal_j,\lch_j|r)$ according to Alg.\,\ref{alg:chdist}.
All code rates use the same channel quantizer.
The quantizer is designed as in Alg.\,\ref{alg:chdes} with input $E_r\{p(\breal_j,\lch_j|r)\}$.
It was observed that each design-$E_b/N_0$ for the rate-compatible decoder had to be slightly higher than the design-$E_b/N_0$ of the corresponding rate-specific decoder. 

We remark that the distributions w.r.t. dimension~$r$ are tracked independently through the node operations in the design phase. 
Individual tracking is useful to compute the average node output distributions as the node degrees depend on the rate $r$.

\subsection{Performance of Rate Compatible Designs}
\cref{fig:rate_compatible_2bit_layer_opt_er} evaluates the rate-compatible design with code rates $\mathcal{R}=\{\frac{1}{3},  \frac{1}{2},\frac{2}{3},\frac{22}{24}\}$ using base graph 1.
The representation in terms of $(\Es/\No)_{\text{dB}}=(\EbNo)_{\text{dB}}+10\log_{10}\myr$ is intentional for better visual separability among the code rates and to highlight the channel conditions where the designed decoders can operate with incremental redundancy (IR) hybrid automatic repeat request (HARQ) techniques~\cite{dahlman23}.

The rate-compatible designs are compared to rate-specific designs with standard and optimized row-layered schedules.
It can be observed that the performance gap compared to the belief propagation decoder is getting smaller for higher code rates.
The rate-compatible designs operate within 0.01\,dB to 0.1\,dB compared to the rate-specific designs.
The 3-bit decoders show up a lower error floor than the 2-bit decoders.
The optimized layered schedule reduces the average iteration count by up to 35\% for the code rate $\nicefrac{1}{3}$.
As mentioned in \cref{sec:layeredopt}, the high irregularity of the lower rate codes makes the schedule optimization more effective.
In some cases, the error floor may increase as cycle effects are not modeled within the layer optimization procedure.

\section{Evaluating Complexity based on Extended State-of-the-Art Hardware Architectures}\label{sec:complexity}
In this section, we discuss the complexity using established decoding architectures from the literature~\cite{lee22multi,jang24area, ren24}.
Many works make use of a row-layered schedule, as it significantly reduces the average number of iterations and can be implemented in hardware with reasonable chip area.
We modify the existing architectures to make better use of coarse quantization to reduce latency and area.
\subsection{Block- and Row-Parallel Decoder Architectures}\label{subsec:block--and-row-parallel-decoder-architectures}
To achieve high area efficiency when decoding long codewords, e.g., with a lifting size of $\Z{=}384$, it is common practice not to execute all parallelizable steps simultaneously in hardware.
Instead, a layer update is divided into multiple computational workloads, each processed in parallel within the constraints of limited hardware resources.
Any remaining parallelism is then exploited by pipelining these workloads.

\cref{fig:block_vs_row_parallel_dec} shows a layer corresponding to a row $i$ in~\cref{fig:orthogonal_rows_cols}.
After lifting the base matrix, one layer consists of $\Z$ rows, each representing a \gls{CN}.
Thus, updating a layer requires computing the \gls{CN} messages for each non-zero entry.

\begin{figure}[t]
	\centering
	\includegraphics[width=0.99\columnwidth]{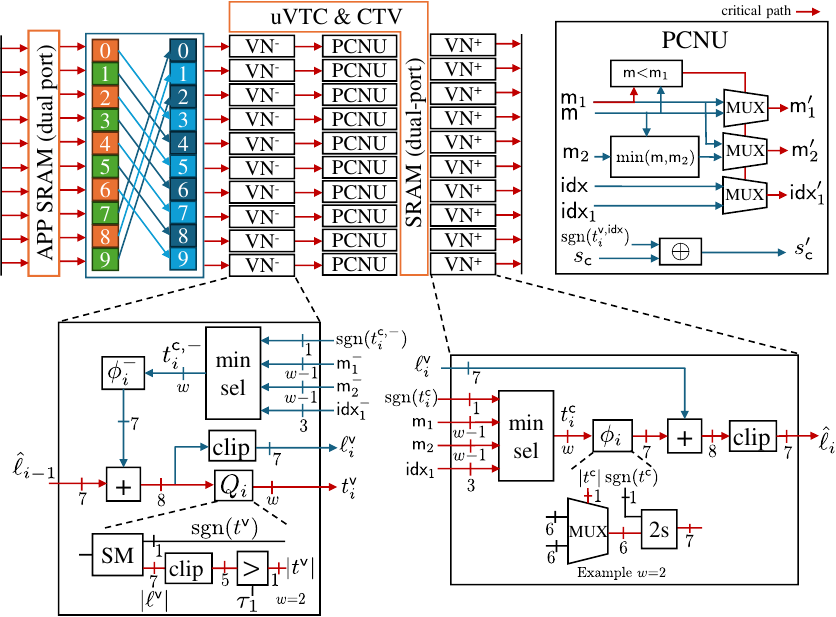}
	\vspace{-0.2cm}
	\caption{A decoder architecture (\acrshort{BAPP}) with block-parallel processing and shifting of \gls{APP} messages (shifts relative to the previous layer)~\cite{ren24}.
		In each layer the inputs to a \gls{CN} with degree $\dc_\idxi$ are processed one after another.
		Thus, $\dc_\idxi$ block updates are required for layer $\idxi$.
		In each block update, the partial \gls{CN} updates the first minimum $\mone$ and second minimum $\mtwo$ using the new incoming message with magnitude $\mmag{=}|\tv|$ at the $\idx$'th \gls{CN} input.
		The illustration uses $\Z{=}10$.
		The 5G decoder uses $\Z{=}384$.}
	\vspace{-0.0cm}
	\label{fig:blockparalleldec}
\end{figure}

Block-parallel processing reduces chip area by computing only $\Z$ \gls{CN} messages simultaneously.
To complete a layer update, $\dc_\idxi$ blocks are processed one after another.
The high-level implementation of a block-parallel decoder is shown in \cref{fig:blockparalleldec}~\cite{ren24}.
A block update loads $\Z$ APP messages $\hatl_{\idxi{-}1}$ from static random-access memory (SRAM), which are permuted relative to the previous layer $\idxi{-}1$ using a shift network~\cite{chen10qsn}.
Then, the $\VNminus$ operation generates a compressed \gls{VN} message $\tv_\idxi$ by subtracting a reconstructed message $\phiminus(\tcminus_\idxi)$ (of the previous iteration) from the APP LLR $\hatl_\idxi$ as shown in \cref{fig:blockparalleldec}, leading to an unquantized \gls{VN} message $\lv_\idxi$.
The reconstruction is done using a lookup table with the input $|\tcminus_\idxi|$ followed by a two's complement (2s) conversion (see~\cite{mohr2023polar} for details).
We convert $\lv_\idxi$ into a sign-magnitude (SM) format.
The magnitude is clipped to 5 bits before being quantized with $\quant_\idxi$.
The partial \gls{CN} update computes the first and second minima by processing all $\dc_\idxi$ \gls{CN} inputs iteratively, using a structure shown in \cref{fig:blockparalleldec}.
In a second phase, after all $\dc_\idxi$ block updates, the $\VNplus$ stage updates the APP message according to \cref{fig:blockparalleldec} and stores it in memory.
Additional memory is required to preserve \gls{UVTC} messages from phase 1 in uVTC SRAM for usage in phase 2~\cite{ren24}.

\begin{figure}[t]
	\centering
	\includegraphics[width=.99\columnwidth]{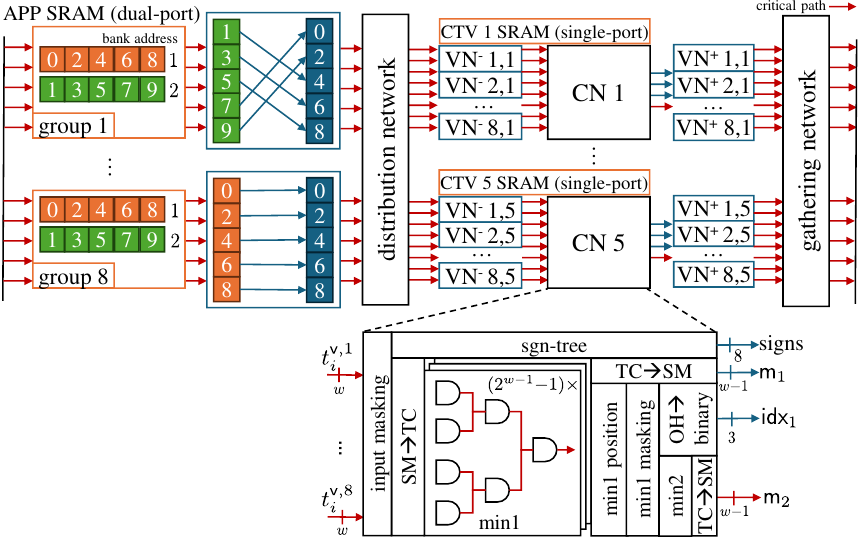}
	\vspace{-0.15cm}
	\caption{A decoder architecture (R-APP) with partial row-parallel processing and shifting of APP messages (shifts relative to the previous layer)~\cite{lee22multi,jang24area}.
		The \gls{VN} updates are as in \cref{fig:blockparalleldec}.
		However, the \gls{CN} update computes all outputs at the same time.
		As a result multiple rows of a decoding layer are updated in parallel.
		The illustration uses parallelism $\Zp{=}5$, lifting $\Z{=}10$ and \gls{CN} input count $\dcR{=}8$. The 5G decoder uses $\Zp{=}32$, $\Z{=}384$ and  $\dcR{=}27$.}
	\vspace{-0.0cm}
	\label{fig:row_app_architecture}
\end{figure}

Another common architecture is the partial row-parallel processing whose computation blocks are shown in \cref{fig:row_app_architecture}.
Large chip area is avoided by updating only $\Zp$ out of $\Z$ rows of a layer in parallel.
In this way, the size of the shift network can be reduced as depicted in \cref{fig:block_vs_row_parallel_dec}.
E.g. in 5G NR, if $\Zp{=}32$ and $\Z{=}384$ only every 12th APP messages is shifted to the target \gls{CN}.
This reduces the delay from $20$ to $12$ compared to a block-parallel shifting unit with $\Zp{=}384$ according to the table in \cref{fig:block_vs_row_parallel_dec}.
We remark that distribution and gathering networks are required, but they have a regular structure and use only 1--2\% of the decoder area in~\cite{jang24area}.

The decoding throughput for block- and row-parallel processing for one iteration is
\begin{align}
	\TPblk{\propto}\frac{\myK}{\Ndblk|\setNr|} \text{ and } \TProw{\propto}\frac{\myK}{\Ndrow\lceil\frac{\Z}{\Zp}\rceil \Nlayersr}.\label{eq:throughput}
\end{align}
We approximate that the critical path delay $\Nd$ in terms of two-input gates is proportional to the inverse of the maximum clock frequency.
The number of block updates $\setNr$ and layers $\Nlayersr$ depends on the code rate $\myr$.

\subsection{Proposed Row-Parallel Architecture}
\begin{figure}[t]
	\centering
	\includegraphics[width=0.99\columnwidth]{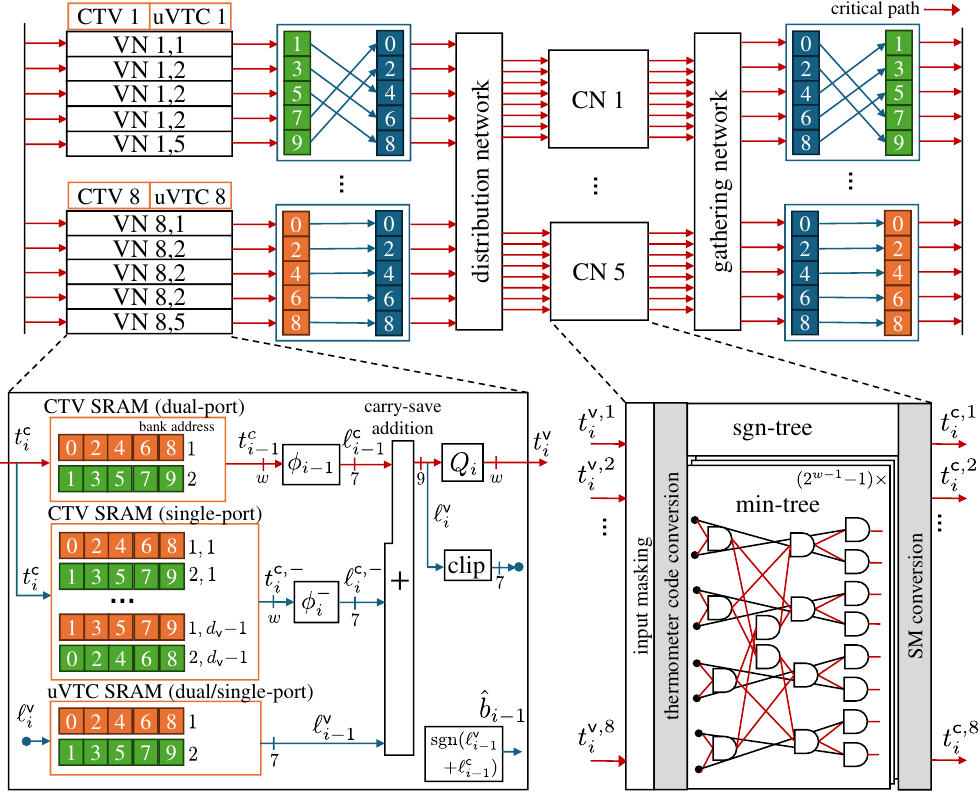}
	\vspace{-0.15cm}
	\caption{Proposed partial row-parallel decoder with permutation of \gls{VN} and \gls{CN} messages (\acrshort{RVC}).
		The \gls{VN} update (b) has shorter critical path by avoiding computation of the APP message.
		The \gls{CN} update (c) uses a thermometer encoding to perform the magnitude computations in a short-delay AND-tree structure.
		The thermometer code encoding and decoding is not required under 2-bit decoding as in \cite{mohr22aware}.}
	\vspace{-0.0cm}
	\label{fig:vcdecoder}
\end{figure}

This section introduces a reorganization of the row-parallel architecture to optimize complexity, particularly targeting improvements in memory demand and the shift network through coarse quantization.
As shown in \cref{fig:vcdecoder}, the shift network is repositioned between the \gls{VN} and \gls{CN} processing stages.

The use of low-resolution messages directly reduces the area of the shift network at the expense of a longer delay, since there is a forward and a backward shift network.
However, this additional delay is compensated by using a faster min-tree at the \gls{CN}, which avoids serial computation of the first minimum $\mone$, its input index $\idxone$, and the second minimum $\mtwo$~\cite{mohr22aware}.
The faster min-tree is enabled by storing the \gls{CN} magnitudes in an uncompressed format $(|t^{\labelcn,1}_i|,\ldots, |t^{\labelcn,\dc}_i|)$ instead of a compressed format $(\mone, \mtwo, \idxone)$.
Further, this avoids delay and area for a minimum selection operation.
Moreover, the rearrangement allows to replace two consecutive additions with one carry-save addition in the \gls{VN}.
We consider a Ladner-Fisher tree-adder for these additions~\cite{koren2001computer}.

\begin{figure}[t]
	\centering
	\includegraphics[width=0.95\columnwidth]{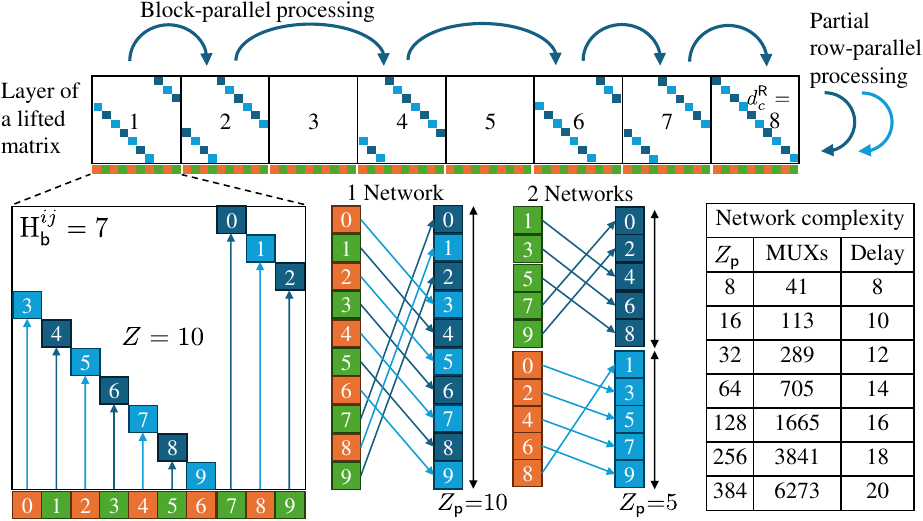}
	\caption{The block-parallel processing updates $\dc_i{=}6$ blocks serially within locations 1 to 8 of row-layer $\idxi$.
		Each block of $\Z$ messages is updated in parallel which requires one large shift network of size $\Zp{=}\Z{=}10$ with comparatively high delay.
		In contrast, the partial row-parallel processing updates $\Zp{=}5$ rows of layer $\idxi$ in parallel, demanding a shift network of size $\Zp{=}5$ with lower delay for each of the $\dcR{=}8$ base columns.
		Only every second row is updated in parallel.
		Thus, \gls{VN} messages must be loaded from every second memory location with an address offset that depends on the cyclic shift value.
		For example, messages from location 1 are loaded with an offset of 1 because the shift value is odd, while the messages from location 2 require no offset because the shift value is even.
		In the 5G decoder $\Zp{=}32$, $\Z{=}384$ and $\dcR{=}27$. Although the base matrix has 68 columns, only the first 26 columns have weight larger one. From the remaining columns $\idxj{\in}\{27,{\ldots},68\}$ at most one is involved in each layer update as they have weight $\dv_\idxj{=}1$.}
	\label{fig:block_vs_row_parallel_dec}
\end{figure}

Note that $\Zp$  \gls{CN} messages $\tc_{\idxi}$ are written to memory with a specific memory address offset (see description of \cref{fig:block_vs_row_parallel_dec}).
However, the $\Zp$ \gls{CN} message $\tc_{\idxi-1}$ of the previous layer update might need to be read with a different offset as the cyclic shifts change between layers.
This holds also for the LLR messages $\lv_\idxi$ and $\lv_{\idxi{-}1}$.
Thus, dual-port SRAM is required to write and read at the same time and allowing different memory offsets.
An exception are degree-one columns, where $\lv_\idxi$ is the channel message such that single-port SRAM suffices.

On the other hand, when reading the old \gls{CN} message $t^{\labelcn,-}_\idxi$, the same memory offset is required as writing $t^{\labelcn}_\idxi$ in the previous iteration.
Hence, reading and writing at the same time can be realized with single-port SRAM as in~\cite{jang24area}.

\subsection{Complexity Comparison of Architectures}
\begin{table}[t]
	\centering
	\caption{Complexity using different architectures and bit widths}
	\label{tab:complexity}
	\vspace{-0.2cm}
	\includestandalone[mode=\mymodestandalone]{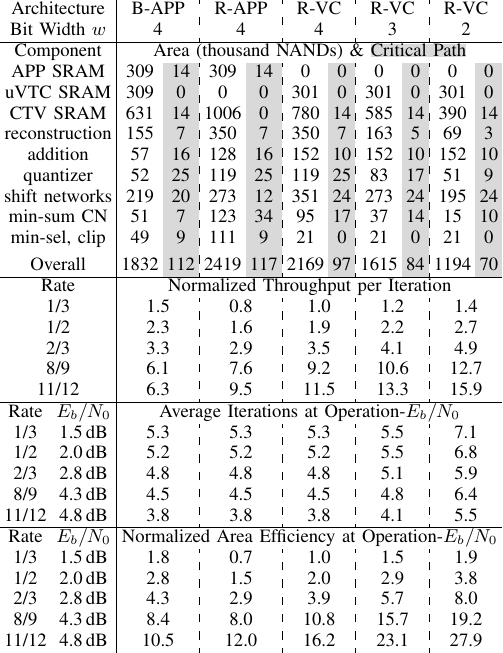}
\end{table}

\begin{table}[t]
	\caption{Area in Terms of CMOS transistor count \cite{gajda2007reducing}} 
	\vspace{-0.2cm}
	\centering
	\includestandalone[mode=\mymodestandalone]{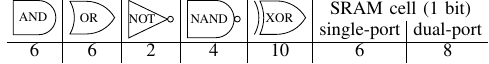}
	\label{tab:complexity_components}
\end{table}
To estimate the chip area, we count the number of logic gates needed for the decoding operations. 
\cref{tab:complexity_components} allows us to convert different types of logic gates into an equivalent number of NAND gates $\Ag$ based on their transistor count.
In addition, we have accumulated the number $\Nd$ of consecutive two-input logic gates on the critical path of the decoder.

In hardware implementations, pipeline stages are commonly used to divide the critical path.
However, since the increase in clock frequency is expected to be approximately proportional across the architectures being compared, we exclude pipeline stages from the evaluation.

Gate count and delay are compared in \cref{tab:complexity} for the \gls{BAPP} decoder, the \gls{RAPP} decoder, and the \gls{RVC} decoder.
The 2-bit VC decoder reduces the area by 45\,\% compared to the 4-bit VC decoder.
Also the delay is significantly reduced, due to faster reconstruction, quantization and \gls{CN} operations. %

\begin{figure}[t]
	\centering
	\includestandalone[mode=\mymodestandalone]{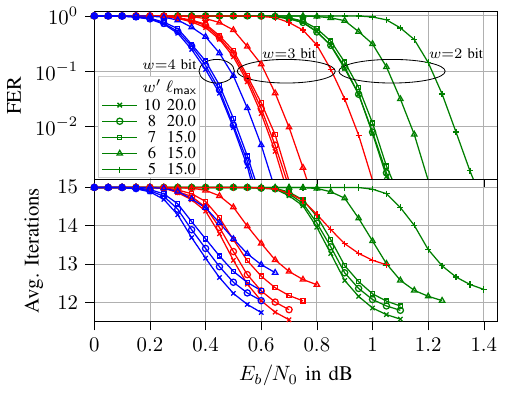}
	\vspace{-.4cm}
	\caption{Performance with different internal adder resolutions. The resolution is $\kappav{=}\lmax/2^{w'{-}1}$ where LLR levels beyond $\lmax$ are clipped. }
	\vspace{-.0cm}
	\label{fig:compare_row_layered_sweep_internal_resolution}
\end{figure}

For all decoders, the memory complexity consumes a significant fraction of the total area, ranging from 50\,\% for the \gls{RVC} decoder to almost 70\,\% for the \gls{BAPP} decoder.
also observed in \cite{jang24area}\cite{ren24}.
From \cref{fig:compare_row_layered_sweep_internal_resolution} we can observe that an internal resolution $w'{=}7$ for the adder units is sufficient to avoid significant loss.

The throughput is calculated with \cref{eq:throughput}.
All normalized throughputs are relative to the 4-bit \gls{RVC} decoder configuration at rate $\nicefrac{1}{3}$.
The evaluation of throughput using \cref{eq:throughput} shows that the 2-bit \gls{RVC} decoder achieves by far the highest peak throughput.
 However, at low code rates, the increased number of layers leads to a substantial throughput reduction under row-parallel decoding.
 In contrast, this reduction is less pronounced for block-parallel decoding, as the number of block updates per decoding layer corresponds to $\dc_\idxi$ which is small for the rows $\idxi$ associated with low code rates (cf. \cref{fig:ratecompatiblebg1bg2})~\cite{ren24}.

\begin{figure}
	\centering
	\includestandalone[mode=\mymodestandalone]{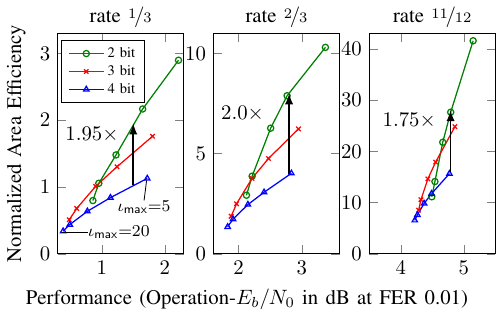}
			\vspace{-0.3cm}
	\caption{Normalized area efficiency compared to FER performance w.r.t. the 4-bit \gls{RVC} decoder ($r{=}\nicefrac{1}{3}$).
	We set the maximum number of iterations as $\itermax{\in}\{20, 15, 10, 7.5, 5\}$ to obtain different operating points, where $0.5$ iterations correspond to row-layer updates updating half of memory locations ($\myK{=}8448$, computational domain \gls{VN}, min-sum \gls{CN}, \gls{CN}-aware design with row alignment, $\wch=5$). }
	\vspace{-0.2cm}
	\label{fig:aevsperf}
\end{figure}

The area efficiency, i.e., throughput divided by the chip area, is calculated as $\AEmy\propto \iteravg\TP/\Ag$.
The average number of iterations $\iteravg$ depends on the operation-$\EbNo$ and the maximum number of iterations $\itermax$.
For every architecture $\itermax$ is chosen, such that the FER equals $10^{-2}$.
In \cref{fig:aevsperf} we evaluate the normalized area efficiency for different maximum number of iterations $\itermax$.
It can be observed that the area-efficiency improves significantly when decreasing $\itermax$ at the expense of SNR loss.
Hardware implementation typically use a maximum number of iterations $\itermax$ close to 5, to minimize decoding latency and maximize energy efficiency~\cite{ren24}.
In this very low iteration regime, 2-bit decoding offers significantly higher area efficiency.

\vspace{-0.15cm}
\subsection{Further Potential for Complexity Reduction}
By exploiting orthogonal rows, the number of layers reduces from 46 to 32 in \cref{fig:orthogonal_rows_cols}.
In turn the throughput of the \gls{RVC} decoder would increase for rate $\nicefrac{1}{3}$ by factor $46/32{=}1.5$.
This gain can be accomplished by duplicating the \gls{CN} processing units in \cref{fig:vcdecoder}.
As the \gls{CN} processing takes only minor area this extension seems worthwhile.
Furthermore, due to space constraints, we have not investigated the impact of the optimized decoding schedule (\cref{sec:layeredopt}) on decoder complexity, which might increase throughput by up to 35\%.
\cref{fig:mi_evolution_degree} shows that memory locations associated with \gls{CN} degree $d_\labelcn {=}19$ provide high mutual information gains only in late iterations.
In contrast, messages from lower degree \glspl{CN} show smaller mutual information gains in late iterations.
This observation suggests that the size of the CTV memory could be reduced by dynamically reallocating memory to different locations depending on iteration $\iota(k)$.

\section{Conclusions}\label{sec:conclusions}
\begin{figure}[t]
	\centering
	\includestandalone[mode=\mymodestandalone]{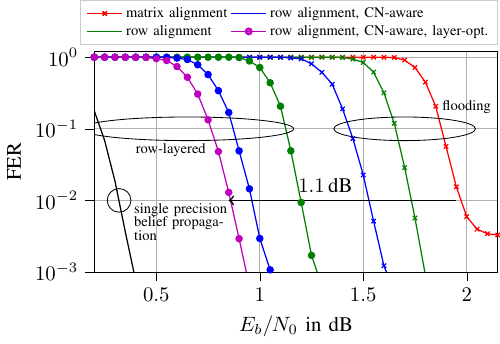}
	\vspace{-0.4cm}
	\caption{Improvements for 2-bit decoding ($\itermax{=}15$, $r{=}\nicefrac{1}{3},\myK{=}8448$).}
	\vspace{0.0cm}
	\label{fig:overall_improvements}
\end{figure}
This paper proposed several techniques for improving low-resolution decoding of 5G-LDPC codes.
First, the base matrix was divided into alignment regions, allowing distinct alphabets of reliability levels for the exchanged messages.
These alphabets were optimized using the \acrlong{IB} method to maximize preserved mutual information.
Employing distinct regions led to performance gains of up to 0.4\,dB under very coarse quantization on graphs with a wide range of node degrees.
Furthermore, the introduced concept of alignment regions enabled a straightforward design of a rate-compatible decoder.
Second, a \gls{CN}-aware design of quantizer thresholds at the \glspl{VN} further improved performance, achieving an additional gain of 0.2\,dB.
Third, the order of layers in row- and column-layered schedules was optimized to prioritize updates that maximize mutual information.
This optimization resulted in throughput increases of up to 35\%.
In summary, selecting appropriate schedules and reliability levels is essential for realizing gains of up to 1.1\,dB under very coarse quantization, as shown in \cref{fig:overall_improvements}.
Finally, a row-parallel decoder architecture was adapted for coarse quantization.
The analysis demonstrated that a 2-bit resolution can double area efficiency compared to a 4-bit resolution while maintaining similar performance.

\bibliographystyle{IEEEtran}
\bibliography{main}
\end{document}